\documentclass[fleqn,usenatbib]{mnras}

\usepackage{newtxtext,newtxmath}

\usepackage[T1]{fontenc}

\DeclareRobustCommand{\VAN}[3]{#2}
\let\VANthebibliography\thebibliography
\def\thebibliography{\DeclareRobustCommand{\VAN}[3]{##3}\VANthebibliography}

\usepackage{graphicx}	
\usepackage{amsmath}	
\usepackage{amssymb}	
\usepackage[dvipsnames]{xcolor}

\newcommand{\vlos}{v_{\mathrm{los}}}
\newcommand{\parallax}{\varpi}

\newcommand{\masyr}{ \ {\rm{mas \ yr^{-1}}}\>}
\newcommand{\mas}{ \ {\rm{mas}}\>}

\newcommand{\kms}{ \ {\rm{km \ s^{-1}}}\>}

\newcommand{\foo}[1]{}

\newcommand{\kpc}{\>{\rm kpc}}

\newcommand{\degree}{\degr}

\newcommand{\hyperfootnote}[1][]{\def\ArgI\hyperfootnoteRelay}
\newcommand\hyperfootnoteRelay[2][]{\href{#1#2}{\ArgI}\footnote{\href{#1#2}{#2}}}

\title[
Velocity gradients in the MW dSphs]{Tidally induced velocity gradients in the Milky Way dwarf spheroidal satellites}

\author[A. M. Martínez-García et al.]{
Alberto Manuel Martínez-García,$^{1,2}$\thanks{E-mail: ammtnez@iac.es}
Andrés del Pino,$^{3}$
and Antonio Aparicio$^{1,2}$\
\\
$^{1}$Instituto de Astrofísica de Canarias, Calle Vía Láctea S/N, E-38205 La Laguna, Tenerife, Spain\\
$^{2}$Universidad de La Laguna, Dpto. Astrofísica, Avda. Astrofísico Fco. Sánchez S/N, E-38206 La Laguna, Tenerife, Spain\\
$^{3}$Centro de Estudios de F\'isica del Cosmos de Arag\'on (CEFCA), Unidad Asociada al CSIC, Plaza San Juan 1, E-44001, Teruel, Spain\\
}

\date{Accepted: 2022 November 8; Revised: 2022 November 8; Received: 2022 June 13}

\pubyear{2021}

\begin{document}
\label{firstpage}
\pagerange{\pageref{firstpage}--\pageref{lastpage}}
\maketitle

\begin{abstract}
We present a kinematic study of six dwarf spheroidal galaxies (dSph) satellites of the Milky Way (MW), namely Carina, Draco, Fornax, Sculptor, Sextans, and Ursa Minor. We combine proper motions (PMs) from the \textit{Gaia} Data Release 3 (DR3) and line-of-sight velocities ($\vlos$) from the literature to derive their 3D internal kinematics and to study the presence of internal velocity gradients. We find velocity gradients along the line-of-sight for Carina, Draco, Fornax, and Ursa Minor, at $\geq 1\sigma$ level of significance . The value of such gradients appears to be related to the orbital history of the dwarfs, indicating that the interaction with the MW is causing them. Dwarfs that are close to the MW and moving towards their orbits pericentres show, on average, larger velocity gradients. On the other hand, dwarfs that have recently left their orbits pericentres show no significant gradients. Lastly, dwarfs located at large Galactocentric distances show gradients with an intermediate intensity. Our results would indicate that the torque caused by the strong tidal forces exerted by the MW induces a strong velocity gradient when the dwarfs approach their orbits pericentres. During the pericentre passage, the rapid change in the forces direction would disrupt such gradient, which may steadily recover as the galaxies recede. We assess our findings by analysing dwarfs satellites from the TNG50 simulation. We find a significant increase in the intensity of the detected gradients as the satellites approach their pericentre, followed by a sharp drop as they abandon it, supporting our results for the dSphs of the MW.
\end{abstract}

\begin{keywords}
galaxies: dwarf--galaxies: evolution  -- galaxies: kinematics and dynamics -- Local Group
\end{keywords}



\section{Introduction}

Dwarf galaxies are considered to be a cornerstone in the galaxy evolution paradigm. In the currently most accepted scenario, the $\Lambda$ Cold Dark Matter ($\Lambda$CDM) model, dwarfs are the first structures to form. Larger galaxies are thus formed through mergers of these dwarf galaxies and gas accretion (\citealt{WhiteRees1978, Blumenthal1984, Dekel1986, NavarroFrenkWhite1995}). Therefore, the dwarfs we observe today would be survivors that have not yet merged into a larger galaxy and could contain key information about the primeval Universe.

The Local Group (LG) offers a unique opportunity to study dwarf galaxies. The proximity of its members allow us to resolve their individual stars, which has opened the door for detailed studies of their star formation histories (\citealt{Aparicio2001, Carrera2002, Monelli2010a, Monelli2010b, delPino2013,deBoer2014, Bettinelli2018, Bettinelli2019}), the chemical abundances of their stars (\citealt{Battaglia2006, Battaglia2008a, Battaglia2011, Koch2006a, Koch2006b, Koch2007b, Kirby2010, Kirby2013, Kirby2015,  Walker2015, Pace2020, Pace2021}), and their internal kinematics (\citealt{Battaglia2008b, Battaglia2011, delPino2017, delPino2017b, Pace2020, delPino2021, MartinezGarcia2021, Battaglia2022b}). The most numerous among the LG are the dwarf spheroidal galaxies (dSphs), systems characterized by their spheroidal shape, lack of gas  (\citealt{Gallagher1994, McConnachie2012}), and their lack of recent star formation (\citealt{Gallagher1994, Aparicio2001, Carrera2002, Bettinelli2018, Bettinelli2019}). From a cosmological point of view, these systems are likely to be an evolved state of late-type dwarfs transformed into gas-poor spheroids due to ram pressure, tidal stirring, and other environmental mechanisms (\citealt{Mayer2010} and references therein). These mechanisms would not only have removed the gas and quenched the star formation, but also disrupted their initial internal kinematics. This scenario, however, faces some difficulties since no significant evidence of internal rotation has been found in the isolated dSphs Cetus and Tucana (\citealt{Taibi2018, Taibi2020}). These galaxies would have had a limited interaction with the Milky Way (MW) or M31 (with at most a pericentric passage), therefore they would be expected to have retained a significant part of their initial rotation. Additionally, the rotation velocity to velocity dispersion ratio ($v_{rot} / \sigma$) does not show significant differences between satellites and isolated dwarfs in the LG (\citealt{Wheeler2017}).
Moreover, should this scenario be correct, some residual traces from the past kinematics should remain even in the internal kinematics of the dSph satellites of the MW (\citealt{Kazantzidis2011, Lokas2015}), which are known to have interacted with their host.

During the last decades, catalogues containing line-of-sight velocities ($\vlos$) for hundreds, or even thousands of stars in nearby dSphs have become widely available. These have allowed the study of the internal kinematics of dSphs, providing key information about their mass profile, mass-to-light ratios and also their systemic motions. Most of these works did not find clear evidence of coherent motions within dSphs, which would explain their spheroidal shape as a consequence of a mostly pressure-supported dynamics (\citealt{Kleyna2002, Wilkinson2004, Munoz2005, Munoz2006, Koch2007a, Koch2007b, Walker2009, Wheeler2017}). On the other hand, some works claim the detection of velocity gradients in the line-of-sight in Sculptor, Sextans, Fornax, and Ursa Minor (\citealt{Battaglia2008b, Battaglia2011, Amorisco2012, Zhu2016, delPino2017, Pace2020}). Whether these gradients are real or rather the projection in the sky of the tangential component of the velocity (\citealt{Feast1961}) has remained a mostly unanswered question due to the lack of reliable proper motions measurements (PMs) for these systems.

The \textit{Gaia} mission (\citealt{GaiaCollaboration2016}) has revolutionized our view of the LG, providing PMs for thousands of stars in nearby galaxies. This allowed us to study the internal kinematics of dSphs in the plane of the sky (\citealt{MartinezGarcia2021}), finding rotation in Carina, Fornax, and Sculptor dSphs. Furthermore, in combination with \textit{Hubble Space Telescope} data, \textit{Gaia} has provided PMs with sufficient precision as to determine 2-dimensional velocity dispersions in some of these systems (\citealt{Massari2018, Massari2020, delPino2022}). The PMs also provide us with the tangential components of the velocity that allow us to correct for the projection of these along the line-of-sight. This opens the door for new studies based on old $\vlos$ data sets, correcting for such perspective effect using \textit{Gaia} PMs. An example of this can be found in \citet{delPino2021}, where the authors found residual rotation in the Sagittarius dSph galaxy after combining \textit{Gaia} PMs with $\vlos$ information from public catalogues.

In this work, we revisit the study of velocity gradients along the line-of-sight for six of the classical dSph satellites of the MW, namely Carina, Draco, Fornax, Sculptor, Sextans, and Ursa Minor. We use the techniques explained in \citet{delPino2021} to combine PMs from the \textit{Gaia} Data Release 3 (DR3; \citealt{GaiaDR3}) and $\vlos$ catalogues from the literature to study the internal kinematics of these systems and detect internal velocity gradients. The paper is organized as follows. In Section~\ref{sec:data_and_methods} we introduce the data and methods. In Section~\ref{sec:results}, we present and discuss the results. Finally, in Section~\ref{sec:conclusions}, we summarize the conclusions of this work.

\section{Data and methods}
\label{sec:data_and_methods}
The study of the internal kinematics of dwarf galaxies requires a solid selection of member stars and the definition of an appropriate reference frame. In this section, we will focus on the stellar membership selection procedure, the internal kinematics derivation, and the technique used to detect the presence of velocity gradients in these systems. 

\subsection{Stellar membership selection}
\label{sec:seleccion}
First, we selected sources from the \textit{Gaia} DR3 data base. We made use of the software \textsc{GetGaia}\footnote{https://github.com/AndresdPM/GetGaia} \citep{delPino2021, GetGaia} for downloading data, selecting members, and discarding poor measurements. In particular, we limited our selection to the quality cuts that are fully explained in Section 2.1.1 of \citet{MartinezGarcia2021} which will be summarized here. In short, we download from the  data base all the stars located within $r \leq 1.5 r_t$ of each galaxy (taking $r_t$ values from \citealt{Irwin1995}). Afterwards, we only give full consideration to sources with available astrometric solution and colours. Furthermore, we impose some broad cuts in PM and parallax, in order to select stars compatible with the galaxies internal kinematics. $\mu_0 - 2.5 \masyr \leq \mu \leq \mu_0 + 2.5 \masyr$, where $\mu_0$ is the PM in the RA and Dec. directions (\citealt{McConnachie2020b}). In the case of the $\parallax$, we require the stars to have  $-2.5 \mas \leq \parallax \leq 1.5 \mas$. 
Since the astrometry of \textit{Gaia} DR3 is unchanged from \textit{Gaia} Early Data Release 3 (EDR3; \citealt{GaiaEDR3}),
we apply the criteria presented in the \textit{Gaia} EDR3 verification papers to screen out poor measurements and correct parallaxes (\citealt{Lindegren2020b}). Random uncertainties present in EDR3 PMs are known to be underestimated. We correct this by multiplying them by a factor 1.05 and 1.22 for 5-parameter and 6-parameter solutions, respectively  (\citealt{Fabricius2020}). As for the correction of the $G$ magnitudes that had to be applied to \textit{Gaia} EDR3 data (\citealt{Riello2020}), it is not necessary anymore given that data from \textit{Gaia} DR3 already includes it (\citealt{GaiaDR3}). 
Finally, we select stars based on their astrometric and photometric errors, keeping only stars with errors below the typical nominal ones in the EDR3 at $G=21$. Namely, only stars with errors smaller than $1 \masyr$ in PM, $0.7 \mas$ in $\parallax$, 0.01 mag in $\mbox{\tt phot\_g\_mean\_mag}$ and 0.1 mag in $\mbox{\tt phot\_bp\_mean\_mag}$ and $\mbox{\tt phot\_rp\_mean\_mag}$ are conserved.

The line-of-sight component of the velocity, $\vlos$, is needed in order to analyse the 3-dimensional internal kinematics of the galaxies. We investigated the presence of member stars in our sample of galaxies with 3D kinematics in the \textit{Gaia} DR3 data, but found no stars. This was expected: the tips of the red giant branch (RGB) of all our galaxies are located well below the nominal limiting magnitude with \textit{Gaia} $\vlos$ measurements ($G\sim14$), which makes \textit{Gaia} unusable for this particular goal. Hence, we continued by cross matching our initial selection with external spectroscopic catalogues with $\vlos$ measurements. We use several catalogues containing $\vlos$ of individual stars, namely; \citet{Walker2009} for Carina, Sculptor, and Sextans; \citet{Walker2015} for Draco; \citet{Pace2020} for Ursa Minor; and \citet{Pace2021} for Fornax. The cross match allows us to construct the 3D velocity vector of the individual stars, which can be used to further clean our sample from MW foreground stars. We perform such cleaning by selecting members in the PM-$\vlos$ space based on the logarithmic likelihood of belonging to a 3D Gaussian distribution ($3 \sigma$) centred in the galaxy's systemic motion (see \citealt{delPino2021}). This procedure is performed iteratively, using as a first guess the systemic PMs from \citet{MartinezGarcia2021} and systemic $\vlos$ from \citet{McConnachie2020a}.

\begin{figure*}
    \centering
    \includegraphics[scale=0.35]{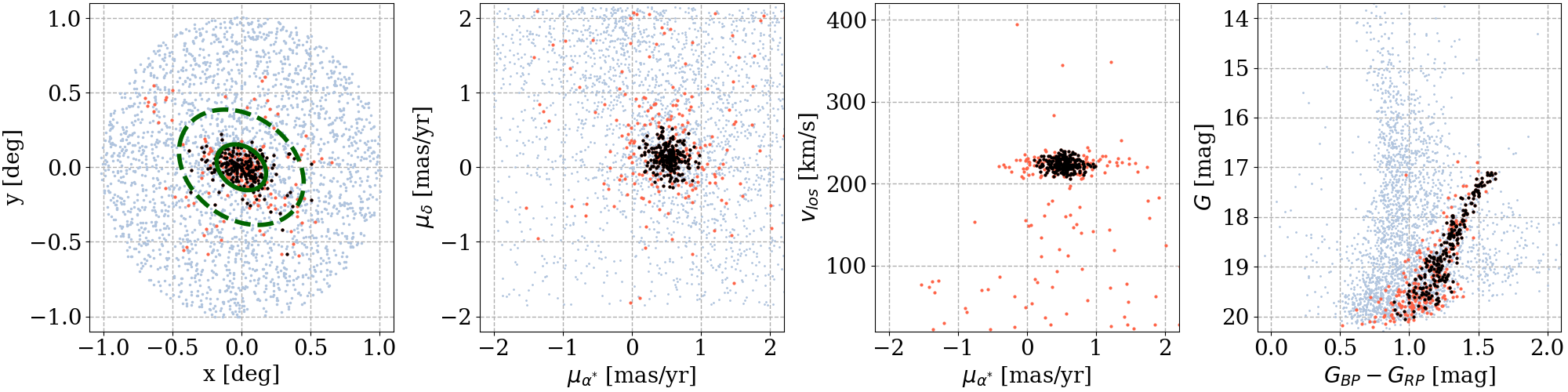}
    \caption{Selection of member stars for Carina. Light blue points represent stars downloaded from the \textit{Gaia} DR3 data base, light red points represent the stars resulting from the cross match, and black points represent the final selection of member stars (see section~\ref{sec:seleccion}). From left to right the plots show the distribution in the sky, in the PM space, in the PM in the RA direction-$\vlos$ space, and in the CMD, respectively. In the leftmost panel, the dashed green ellipse represents the tidal radius (\citealt{Irwin1995}) and the solid green line the half-light radius (\citealt{McConnachie2020a}).}
    \label{fig:selection}
\end{figure*}

In Figure~\ref{fig:selection} we show an example of the selection of sources for Carina, a galaxy for which we obtained a sample of intermediate size (236 members, see Table~\ref{tab:resultadosv}). Results show a consistent member selection across all panels. From the selection in the sky we find that the vast majority of the sources lie within the tidal radius. The selection in the PM-$\vlos$ displays a proper selection of sources, with no obvious outliers. Finally, the selection in the colour-magnitude diagram (CMD) clearly shows that member stars are mostly located in the RGB, as expected  for an intermediate-old stellar population. Some stars that are likely members are rejected by our selection method. These are stars with typically larger astrometric uncertainties, which causes them to lie outside the $3\sigma$ Gaussian distribution that we use for the selection. We performed tests using less restrictive selection thresholds ($4\sigma$, $5\sigma$), obtaining consistent results with those obtained clipping the sample at $3\sigma$. Therefore, we decided to keep our initial, more restrictive, selection at $3\sigma$, which produced cleaner samples of well measured stars (i.e. with low uncertainties in the PMs). 

\subsection{Internal kinematics derivation}
\label{subsec:internal_kinematics}
The study of the internal kinematics requires introducing a co-moving reference frame centred on each galaxy. Its formulation is presented in \citet{VanderMarel2001} and \citet{VanderMarel2002}. Further explanation can be found in \citet{delPino2021} and \citet{MartinezGarcia2021}, where it was specifically applied to dSph satellites. 
Briefly explained, the 3-dimensional systemic motion of the galaxy's centre-of-mass (CM) is subtracted from the sky-projected velocity field. Such systemic motion vector is built using the systemic PMs presented in \citet{MartinezGarcia2021} and the error-weighted average of the $\vlos$ of its member stars. After the subtraction of the CM motion, the 3-dimensional relative velocities of the stars are decomposed into three orthogonal components:

\begin{equation}\label{eq:v1v2v3}
v_{S,i} \equiv \frac{dD_i}{dt},\\ v_{R,i} \equiv D_i\frac{d\rho_i}{dt},\\ v_{T,i} \equiv D_i\sin{\rho_i}\frac{d\phi_i}{dt} ,
\end{equation}  
where $D_i$ is the heliocentric distance, $\rho_i$ is the angular distance to the CM, and $\phi_i$ is the position angle (measured North to East) of the $i$-th star. Therefore, $v_{S,i}$ is the line-of-sight component of the velocity, and $v_{R,i}$, $v_{T,i}$ are the radial and tangential components of the velocity in the plane of the sky, all of them defined with respect to the CM. Note that the reference frame is unique to each galaxy and referred to its CM, so it does not change star by star.
Moreover, the individual velocities are also decomposed in three additional orthogonal components referred to the CM: $v_{x,i}$, $v_{y,i}$, $v_{z,i}$ where $x$-axis is anti parallel to the RA axis; $y$-axis is parallel to the Dec axis; and $z$-axis points towards the observer for a star located at the CM. The equations of the velocity components expressed in terms of directly observable quantities (RA, Dec, distance, PMs, and $\vlos$) can be found in Section 6 of \citet{delPino2021}. Note that in this paper and in \citet{MartinezGarcia2021} the velocity components were renamed for the sake of clarity, so $v_S$, $v_R$, and $v_T$ correspond to $v_1$, $v_2$, and $v_3$ in \citet{delPino2021}, respectively. In Figure~\ref{fig:observation_point} we show the reference frame, the observer, and the velocity components (for further clarification, see Figures 1 and 2 from \citealt{VanderMarel2002}). Given that there are no accurate measurements of distances to individual stars in dSphs, we assume that all the stars of each galaxy are located at the same distance $D_0$ (\citealt{McConnachie2020a}).
Although this is not exact, the assumption has a negligible impact on the velocity components of the stars, introducing differences that are well below the observational errors.

In order to obtain the associated uncertainties of the velocity components, we use a Monte Carlo (MC) scheme that simulates systematic errors and propagates all known random uncertainties to the final results. In short, in each iteration  all the variables used to compute the velocity components of each star ($v_{S,i}$, $v_{R,i}$, $v_{T,i}$, $v_{x,i}$, $v_{y,i}$, $v_{z,i}$) are randomly sampled from a normal distribution centred in the corresponding nominal value of the variable and with standard deviation given by the corresponding random errors (see Section~\ref{sec:seleccion}). The simulation of systematics errors attempts to reproduce the observed small-scale systematic effects reported in EDR3 data. This simulation follows the prescription explained in \citet{VanderMarel2019} and \citet{delPino2021}, although using a smaller amplitude, $56\mu\mathrm{as}$ yr$^{-1}$. This amplitude provides a root-mean-square compatible with the one observed in \textit{Gaia} EDR3 at scales of $\sim 1 \deg$ (see \citealt{Lindegren2020}). We perform  $10^4$ iterations following this procedure, adopting the standard deviation of the values obtained for each star as the total error of the velocity.

\subsection{Detection of gradients in $v_z$}
\label{sec:gradients}
Velocity gradients along the line-of-sight in dSphs provide insightful information about their internal kinematics. We used $v_z$ to study possible velocity gradients in these systems. The main advantage of using $v_z$ is that, unlike $\vlos$, it is not affected by perspective effects which could induce spurious gradients (\citealt{Feast1961}).
In order to detect gradients in $v_z$, we fit a plane to the components of the position $x$, $y$ and velocity $v_z$ of the stars of each galaxy using a Monte Carlo Markov Chain (MCMC) regression model. The plane is defined as follows, $v_z = a_1\,x + a_2\,y + b$, being $a_1$, and $a_2$ the fitting coefficients and $b$ the intercept. The model considers not only the nominal values of the positions and velocities but also their corresponding uncertainties, obtained through the MC simulations described in Section~\ref{subsec:internal_kinematics}. This is crucial for deriving robust estimations of the fitting coefficients, intercept, and intrinsic dispersion of $v_z$, whereby it is possible obtaining the amplitude of the gradient (henceforth denoted as $A_{\mathrm{grad}}^{v_z}$, being the square root of the quadratic sum of the coefficients, $\sqrt{a_1^2 + a_2^2}$) and its direction in the sky 
(derived as the arctan of the coefficients, $\arctan(a_2/a_1)$, and given as position angle, PA, measured North to East).
We define the direction of the gradient as the direction in which it increases, and consider that a galaxy shows a non-negligible gradient when $A_{\mathrm{grad}}^{v_z}$ is not compatible with zero within $1\sigma$.

\subsection{Rotation in the plane of the sky and velocity gradients}
\label{subsec:rotation_grad}
\begin{figure}
    \centering
    \includegraphics[scale=0.15]{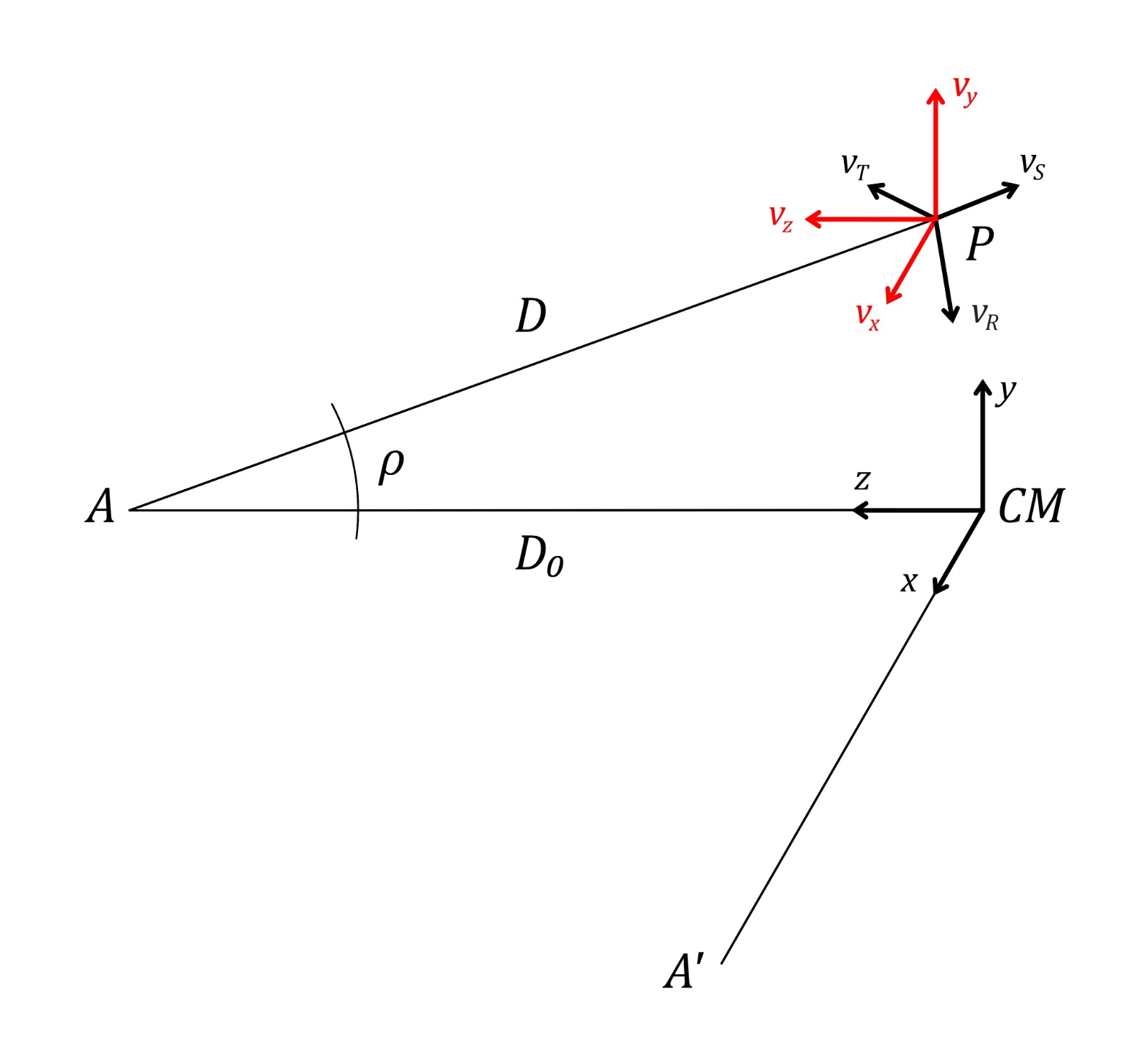}
    \caption{Representation of the reference frame, velocity components, and the observers.  The centre of the orthogonal system ($x$, $y$, $z$) is $CM$, the CM of the galaxy. $A$ represents the position of the observer (Section~\ref{subsec:internal_kinematics}). $D_0$ is the distance from $A$ to the CM of the galaxy. $D$ is the distance from $A$ to a star (located at $P$). 
    $A'$ is the position of the observer after rotating it 90 deg around the $y$-axis (Section~\ref{subsec:rotation_grad}). $y$-$z$ coincides with the plane of the figure, that contains $CM$ and $A$. The rest of elements are in schematic perspective. The definition of all the quantities can be found in Section~\ref{subsec:internal_kinematics}, \citet{VanderMarel2001}, and \citet{  VanderMarel2002}.}
    \label{fig:observation_point}
\end{figure}

Previous studies have derived the rotation velocity in the plane of the sky ($V_T$) for the dSphs that we analyse in this paper (\citealt{MartinezGarcia2021}). The gradients that we derive in  Section~\ref{sec:gradients} could be linked to rotation along the line of sight, i.e. in perpendicular direction to the reported rotation velocities in the plane of the sky.
In order to compare rotation velocities in the plane of the sky to the amplitude of the gradients in $v_z$, it is necessary to operate in an equivalent frame, since velocities and gradients cannot be compared directly. The rotation of a stellar system can be detected as a velocity gradient, therefore it is possible to derive the gradient associated with the reported rotation in the plane of the sky. To measure such gradients, we use a MC scheme. For each iteration, the spatial and velocity coordinates of the stars of a galaxy are sampled from a random normal distribution centred on their corresponding nominal values. This allows the propagation of all known uncertainties. Then we rotate the position of the observer with respect to the galaxy 90 deg around the $y$-axis. In Figure~\ref{fig:observation_point} we show the reference frame, the initial position of the observer ($A$), and its new position ($A'$). The 90 deg rotation of the observer around the $y$-axis allows to translate the rotation in the plane of the sky of a galaxy measured by an observer in $A$ into a velocity gradient along the line of sight for an observer placed in $A'$ by geometric means. For a galaxy rotating in the plane of the sky, an observer placed in $A$ witnesses the coherent motion of the stars around the CM in the $x-y$ plane (i.e. rotation around the $z$-axis). From the point of view of an observer placed in $A'$, the very same rotation is revealed in the plane $y-z$ by a velocity gradient in $v_x$, whose amplitude can be derived following a procedure analogous to the one described in Section~\ref{sec:gradients}. This gradient due to the rotation in the plane of the sky can now be properly compared to those derived in Section~\ref{sec:gradients}. 
This is also true for any observer looking at the CM from a  perpendicular direction to the $z$-axis. 
In order not to privilege any particular orientation of the galaxy with respect to $A'$,
we rotate it around the $z$-axis a hundred different angles ($\theta$) ranging from 0 to $2\pi$ and we measure the amplitude of the gradient due to the rotation in the plane of the sky, $a_{\mathrm{grad}}^{V_T}(\theta)$, from $A'$ for each of them.
For the derivation of $a_{\mathrm{grad}}^{V_T}(\theta)$, we fit only $y$ and $v_x$ since we assumed that $D=D_0$ (Section~\ref{subsec:internal_kinematics}), and therefore the $z$ coordinate does not provide relevant spatial information. 
The mean amplitude of the gradients measured from $A'$ for the different $\theta$ angles is calculated as follows:

\begin{equation}
   \overline{A_{\mathrm{grad}}^{V_T}} = \frac{1}{2\pi} \int_{0}^{2\pi} a_{\mathrm{grad}}^{V_T}(\theta) \, d\theta
\end{equation}

The procedure is repeated in $10^3$ iterations. The amplitude of the gradient produced by the rotation in the plane of the sky ($A_{\mathrm{grad}}^{V_T}$) of a galaxy is given as the mean value of individual mean gradients of the realizations. The adopted uncertainty is the standard deviation of all realizations. Note that since we are fitting $y$ and $v_x$, for a galaxy rotating in the plane of the sky in clockwise direction $A_{\mathrm{grad}}^{V_T} > 0$, and for anticlockwise direction $A_{\mathrm{grad}}^{V_T} < 0$.  In the following, for simplicity, we will sometimes use the term velocity gradients to refer to gradients in $v_z$, given that the discussion will be mainly focused on them. For the gradients due to the rotation in the plane of the sky, we will always refer to them explicitly.

\section{Results and discussion}
\label{sec:results}

\begin{table*}
    \centering
    \caption{Results of this study. Columns: (1) galaxy name, (2) Galactocentric distance, (3) orbital period for the "Light MW" potential of \citet{Battaglia2022}, (4) orbital period for the "Heavy MW" potential of \citet{Battaglia2022} (5) number of selected member stars, (6) direction of the gradient in $v_z$, (7) intercept, (8) velocity dispersion in $v_z$, (9) amplitude of the gradient in $v_z$, (10) amplitude of the gradient generated by the rotation in the plane of the sky, and (11) radial Galactocentric velocity.}
    \setlength{\tabcolsep}{5pt}
	\begin{tabular}{l|r|r|r|r|r|r|r|r|r|r} 
	    \hline
		Galaxy & 
		\multicolumn{1}{c}{$d_{GC}$} &
		\multicolumn{1}{c}{$T_l$} &
		\multicolumn{1}{c}{$T_h$} &
		\multicolumn{1}{c}{$n_{*}$} & \multicolumn{1}{c}{PA} & \multicolumn{1}{c}{$b$} & \multicolumn{1}{c}{$\sigma_{v_z}$} & \multicolumn{1}{c}{$A_{\mathrm{grad}}^{v_z}$} & \multicolumn{1}{c}{$A_{\mathrm{grad}}^{V_T}$} & \multicolumn{1}{c}{$v_R^{GC}$}\\
		& \multicolumn{1}{c}{(kpc)} &
		\multicolumn{1}{c}{(Gyr)} &
		\multicolumn{1}{c}{(Gyr)} &
		& \multicolumn{1}{c}{(deg)} & \multicolumn{1}{c}{(km s$^{-1}$)} & \multicolumn{1}{c}{(km s$^{-1}$)} &\multicolumn{1}{c}{(km s$^{-1}$ kpc$^{-1}$)} & \multicolumn{1}{c}{(km s$^{-1}$ kpc$^{-1}$)} & \multicolumn{1}{c}{(km s$^{-1}$)}\\
		\hline
		\hline

Draco      &   $75.9 \pm 5.9$ & $3.06^{+0.35}_{-0.49}$ &  $1.53^{+0.11}_{-0.17}$ &  274  & $144^{+35}_{-35}$    & $0.18^{+0.48}_{-0.48}$  & $ 7.96^{+0.36}_{-0.33}$ & $5.01^{+3.46}_{-3.46}$ & $ 8.4 \pm 22.6$ & $-103.44 \pm 0.66$ \\ 
Ursa Minor &   $77.8 \pm 3.5$ & $2.47^{+0.14}_{-0.17}$ &  $1.34^{+0.10}_{-0.07}$ & 329  & $327^{+43}_{-44}$    & $-0.07^{+0.41}_{-0.41}$ &  $6.88^{+0.32}_{-0.30}$ & $2.76^{+2.42}_{-2.42}$ & $ 2.5 \pm 19.3$ &  $-81.00 \pm 0.51$ \\
Sculptor   &   $86.1 \pm 5.5$ & $3.46^{+0.46}_{-0.32}$ &  $1.66^{+0.13}_{-0.08}$ &  930  & $127^{+100}_{-100}$  & $0.25^{+0.27}_{-0.27}$  &  $7.69^{+0.20}_{-0.19}$ & $0.76^{+1.18}_{-1.18}$ & $ 4.1 \pm  9.1$ &   $75.88 \pm 0.46$ \\ 
Sextans    &   $89.0 \pm 3.9$ & $6.62^{+1.72}_{-1.69}$ &  $2.21^{+0.24}_{-0.27}$ & 146  & $246^{+91}_{-91}$    & $-0.29^{+0.51}_{-0.51}$ &  $5.59^{+0.40}_{-0.36}$ & $1.08^{+1.58}_{-1.58}$ & $15.0 \pm 21.0$ &   $86.96 \pm 0.72$ \\ 
Carina     &  $106.8 \pm 6.3$ & $7.10^{-}_{-2.92}$ &  $2.58^{+1.28}_{-0.53}$ &  236  & $81^{+49}_{-49}$     & $-0.27^{+0.46}_{-0.46}$ &  $6.16^{+0.36}_{-0.33}$ & $3.01^{+1.83}_{-1.84}$ & $25.2 \pm 23.4$ &    $8.79 \pm 0.67$ \\ 
Fornax     & $149.1 \pm 12.1$ & $4.36^{+1.95}_{-0.77}$ &  $2.50^{+0.41}_{-0.27}$ & 1748  & $153^{+21}_{-21}$    & $-0.14^{+0.28}_{-0.28}$ & $11.02^{+0.20}_{-0.19}$ & $1.30^{+0.45}_{-0.45}$ & $ 4.5 \pm  3.0$ &  $-35.11 \pm 0.31$ \\ 

		\hline
	\end{tabular}
	\label{tab:resultadosv}
\end{table*}

\subsection{Observed velocity gradients}

The results from the fit of the linear model are listed in Table~\ref{tab:resultadosv}. The model simultaneously fits the intercept ($b$), velocity dispersion along the $z$-axis ($\sigma_{v_z}$), and the regression coefficients from which we derive the amplitude ($A_{\mathrm{grad}}^{v_z}$), and the direction of the velocity gradients (given as the position angle, PA, measured North to East).

The model does a good job fitting the observed $v_z$. We find $b$ consistent with zero for all the galaxies in our sample, as expected after subtracting their systemic 3D motion. The values of the velocity dispersion along the $z$ direction, $\sigma_{v_z}$, also provide useful information about the performance of the model if we compare them to the line-of-sight velocity dispersion, $\sigma_{v_{los}}$. 
Note that $\sigma_{v_z}$ is measured in an orthogonal system, whereas $\sigma_{v_{los}}$ is measured along the line of sight, therefore the values are  roughly similar in the inner parts of the galaxy and diverge at large angular distances from the galaxies centres. Given the relatively small angular sizes of the analysed galaxies ($\lesssim 1 \degree$), we expect both measurements to be similar, on average.
From our model we find $\sigma_{v_z}$ values that are fully consistent with $\sigma_{v_{los}}$ for most of our galaxies. Only in the cases of Sextans and Ursa Minor, the results differ at $\sim1.4\sigma$ and $\sim1.8\sigma$ level, respectively (\citealt{Walker2009clean, Walker2009univ}). Note that our procedure is not specifically aimed to derive velocity dispersions, however, our results are generally in good agreement with previous measurements.

\begin{figure}
    \centering
    \includegraphics[scale=0.55]{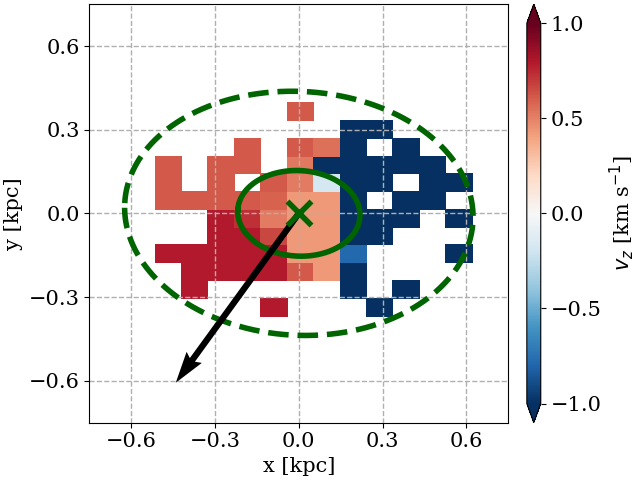}
    \caption{Gradient in $v_z$ for Draco. The centre of the galaxy is marked by a cross. The direction of the gradient is represented by a black arrow. The mean value of $v_z$ within Voronoi cells is marked in the colour bar. The rest of the markers coincide with those of Figure~\ref{fig:selection}.}
    \label{fig:draco}
\end{figure}

\begin{figure*}
    \centering
    \includegraphics[width=\linewidth]{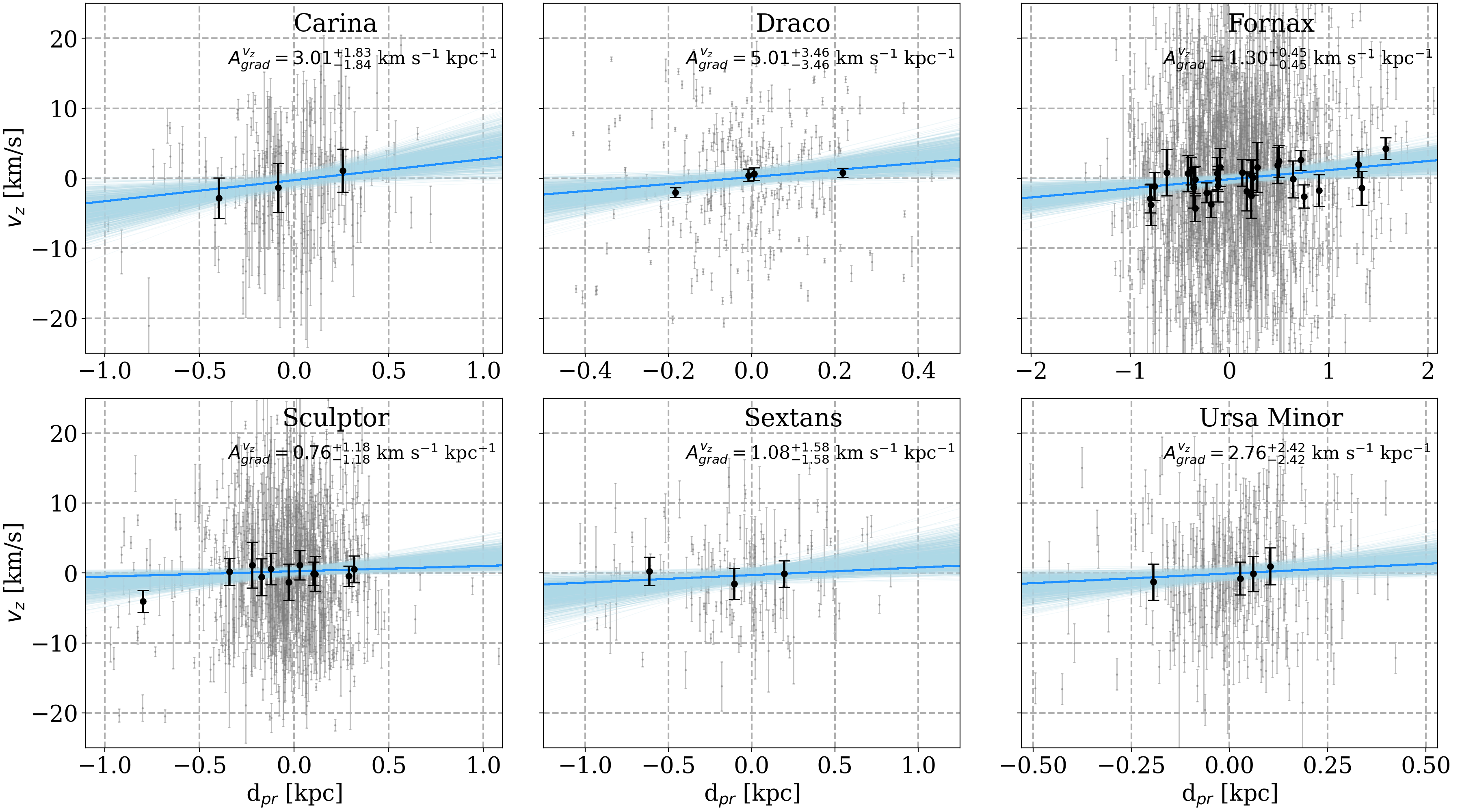}
    \caption{$v_z$ velocities of the individual stars and their projected distance along the direction of the maximum velocity gradient, $d_{pr}$ (with PA listed in Table~\ref{tab:resultadosv}). Grey points represent each individual star, while black points represent the average velocities within Voronoi cells of $\sim50$ stars. Error bars show the corresponding uncertainties. The model that best fits the data through a MCMC is shown by a blue line. Light blue lines are a thousand realizations of the MCMC chosen at random.}
        
    \label{fig:maximo_gradiente}
\end{figure*}

We find velocity gradients in $v_z$ with more than $1 \sigma$ of statistical significance for Carina, Draco, Fornax, and Ursa Minor. On the other hand, only Carina and Fornax show gradients in $v_T$ at more than 1$\sigma$ level of significance. We notice in any case, that the detected gradients are below 
$3\sigma$ level of significance.   
In Figure~\ref{fig:draco} we show the velocity map of $v_z$ for Draco. In order to increase the signal-to-noise, we apply Voronoi tessellation to the data (\citealt{Capellari2003, delPino2017}), generating cells containing $\sim 50$ stars each one. The recovered velocity gradient PA (see Table~\ref{tab:resultadosv}), represented by a black arrow, is consistent with the observed velocity gradient in the Voronoi cells. In Figure~\ref{fig:maximo_gradiente} we show the $v_z$ velocities of the individual stars of each galaxy and their projected distance ($d_{pr}$) to the adopted CM of the galaxy along the direction of the corresponding velocity gradients (Section~\ref{sec:gradients}). Note that $d_{pr}$ is positive for stars which are along the direction in which the gradient increases. The projection of the plane fitted to $v_z$ along this direction is shown by a blue line, with its uncertainty represented by a lighter blue area. 
In order to increase the signal-to-noise ratio and provide a clearer picture of the internal kinematics of the galaxies, we also show the average results within Voronoi cells of $\sim 50$ stars.

\subsection{Comparisons with previous works}

Previous studies have analysed the internal kinematics of these galaxies using $\vlos$. Because of the lack of reliable PMs measurements for most of these galaxies before $Gaia$ DR2, many of these works only correct for the reflex motion of the Sun along the line of sight; i.e. analysing $\vlos$ in the Galactic Standard of Rest (GSR). However, $\vlos$ in GSR are still affected by the systemic transversal motion of the dwarf with respect to the MW which, due to perspective effects, also projects along the line of sight (\citealt{Feast1961}). Therefore, not taking the PMs into account would produce an artificial gradient in the $\vlos$ of the stars along the direction of motion. The effect could also potentially conceal a real gradient or modify its perceived orientation and amplitude. In the following, we offer a compilation of results from previous studies that have analysed the galaxies in our sample. To check for consistency with our results, we try to reproduce their results using our samples of stars. For comparisons with works that did not use PMs, we transformed our $\vlos$ into the GSR instead of using our co-moving reference frame, and proceed analogously to Section~\ref{sec:gradients} to derive the velocity gradients in GSR . Notice that, by definition, a gradient detected by our model in $\vlos$ will have the opposite direction of one detected in $v_z$.

\begin{itemize}
    \item \textit{Carina:} no significant evidence for rotation has been found in the inner parts of the galaxy using $\vlos$ in GSR (\citealt{Munoz2006}). We do not find any significant gradient in $\vlos$ in GSR ($1.24^{+1.84}_{-1.84} \kms\mathrm{kpc}^{-1}$) using our sample of stars for Carina, being consistent with \citet{Munoz2006}. As explained above, $\vlos$ in GSR can be affected by perspective effects, unlike $v_z$, for which shows a clear internal velocity gradient in Carina. \citet{Fabrizio2016} also studied the kinematics of Carina, finding evidence of rotation along the semimajor axis. The study used $\vlos$ and did not correct from perspective effects or transformed the $\vlos$ to GSR. The velocity gradient in $\vlos$ reported by \citet{Fabrizio2016} (PA$\sim 65$ deg) is roughly aligned with the PM of Carina (PA $\sim77$ deg, \citealt{MartinezGarcia2021}), what suggests that the gradient they found could be related to perspective effects.

    \item \textit{Draco:} using $\vlos$, \citet{Kleyna2002} detected a rotation signal of $6 \kms$ about an axis with PA = 62 deg measured at 30 arcmin from the galaxy centre. Given that the north-west part of the galaxy is receding (\citealt{Kleyna2002}) and that the velocity gradient is perpendicular to the rotation axis, this rotation detected in $\vlos$ translates into a gradient in $v_z$ with PA = 152 deg, which is in quite good agreement with the gradient in $v_z$ we observe here ($3.31\kms$ measured at 30 arcmin from the centre, PA = $144^{+35}_{-35}$ deg). Interestingly, the authors of that work did not transformed their $\vlos$ to GSR or tried to correct from perspective effects (although the authors are aware of these effects and provide a thoughtful discussion of their results in this context). The bulk PM vector of the galaxy has a PA $\sim 169$ deg (\citealt{MartinezGarcia2021}), which would introduce a gradient almost in the opposite direction to the one detected by \citet{Kleyna2002}. This would indicate that the reported gradient is real.

    \item \textit{Fornax:} \citet{Walker2008} reported a velocity gradient of $6.3 \pm 0.2$ $\kms$deg$^{-1}$ at PA = 120 deg, using $\vlos$. Using our set of $\vlos$ for Fornax, we reproduce this gradient, obtaining $3.6 \pm 1.2 \kms$ deg$^{-1}$ at PA = $114 \pm 18$ deg. These gradients in $\vlos$ are likely to be induced by projection effects, since both point in the direction of the bulk PM of Fornax (PA = 134 deg, \citealt{MartinezGarcia2021}).
    Chemo-dynamical studies have revealed complex rotation patterns depending on the metallicity of the stars of Fornax (\citealt{delPino2017}). The main rotation signal of Fornax has been measured as a gradient in $v_z$ with PA $\sim 210$ deg, roughly along its semimajor axis (\citealt{delPino2017}). The reported gradient is unlikely to be due to perspective effects since it is almost in perpendicular direction to the bulk PM of the galaxy in GSR (PA $= 134 \pm 1$ deg). Indeed using the sample of stars of \citet{delPino2017}, we proceed as in Section~\ref{sec:gradients} and we obtain a gradient in $v_z$ of $A_{\mathrm{grad}}^{v_z} =0.70 \pm 0.42\kms$kpc$^{-1}$ at PA = $193 \pm 41$ deg, consistent with the main rotation pattern shown in  fig 13 of \citet{delPino2017} (for metal poor stars). The gradient in $v_z$ that we find in this work is pointing in a slightly different direction, PA = $153^{+21}_{-21}$ deg (although is compatible within the error bars).  Such difference is likely to be due to the different samples used in both studies. 
    
    \item \textit{Ursa Minor:}  hints of prolate rotation of $\sim2\kms$ have been reported, with the peak amplitude taking place close to the semiminor axis ($\theta = 140 \degr, 320 \degr$, \citealt{Pace2020}). The authors of that study have taken into account possible perspective effects. 
    The direction of the reported velocity gradient (PA = 140 deg), is equivalent to a gradient in $v_z$ with PA = 320 deg (a gradient in $\vlos$ will have the opposite direction to one detected in $v_z$), which is in agreement with the velocity gradient that we find for Ursa Minor (PA = $327^{+43}_{-44}$ deg).
    
\end{itemize}

We do not find gradients in $v_z$ for Sculptor or Sextans. The absence of gradients in $v_z$ in these galaxies suggests that they are not experiencing rotation along the line-of-sight direction. However, previous studies have reported the presence of $\vlos$ gradients in these two galaxies. 

\begin{itemize}
    \item \textit{Sculptor:} multiple studies have studied the internal kinematics of Sculptor. Using $\vlos$, \citet{Walker2008} reported a velocity gradient of $5.5 \pm 0.5 \kms\mathrm{deg}^{-1}$ at PA = 201 deg. Using our sample of stars for Sculptor we detect a non-significant gradient of $1.13^{+2.32}_{-2.63}\kms\mathrm{deg}^{-1}$ at PA = $204^{+93}_{-94}$ deg, finding that the amplitude is not compatible with the one reported by \citet{Walker2008}, yet the direction of the gradients (despite the large uncertainties) are in good agreement. The differences among the gradients are likely due to the different samples used in both studies. 
     \citet{Battaglia2008b} reported a velocity gradient  of $7.6^{+3.0}_{-2.2}\kms\mathrm{deg}^{-1}$ with PA $\sim 279$ deg (along the projected major axis of the galaxy), using $\vlos$ in GSR. With our sample of stars for Sculptor, we detect a $\vlos$ gradient in GSR of $4.3^{+1.9}_{-1.9} \kms\mathrm{deg}^{-1}$ at PA=$318^{+25}_{-26}$ deg.
     This gradient and the one reported by \citet{Battaglia2008b} lie in the same quadrant where the direction of motion of the galaxy in GSR points (PA = $320\pm2$ deg), which suggests that the projection of the systemic motion along the line of sight could be, at least partially, behind this velocity gradient. Interestingly, the preferred direction of the gradient that we detect in $v_z$ for Sculptor (despite the large uncertainties due to its low significance), PA = $127^{+100}_{-100}$ deg, when expressed in the frame of $\vlos$ (PA = $307^{+100}_{-100}$ deg , note that by definition $v_z$ and $\vlos$ point in opposite directions) is similar to that of the one found by \citet{Battaglia2008b}, which could indicate that some weak internal gradient could exist along the projected major axis of the galaxy.
    Chemo-dynamic studies have reported a weak rotation signal in the metal-rich population of Sculptor along the projected major axis (\citealt{Zhu2016}). The study has taken into account possible perspective effects however, the systemic PM used for the correction are not in agreement with the PM in Dec direction that we use in this study (\citealt{MartinezGarcia2021}) nor with other PMs in Dec direction derived using the \textit{Hubble Space Telescope} or \textit{Gaia} EDR3 (\citealt{Sohn2017, McConnachie2020b, Li2021, Vitral2021, Battaglia2022, Pace2022, Qi2022}).
    Additionally, we note that in our study we do not separate the stars of the sample in different populations, rather we study the overall gradient of the galaxy, which could also explain these discrepancies.
    
    \item \textit{Sextans:} two gradients in $\vlos$ in GSR have been reported in this galaxy (\citealt{Battaglia2011}); one along the semimajor axis (PA = $56$ deg)  of $8.5 \pm 3.0 \kms\mathrm{deg}^{-1}$, and another one along PA = $11$ deg of $7.5^{+3.4}_{-3.0} \kms\mathrm{deg}^{-1}$. 
    That particular study is capable of detecting multiple gradients because it uses slits oriented in several directions and derives the gradient of the stars within the slit. On the contrary, our method is aimed to obtain the overall gradient of the galaxy. We tried to reproduce these results using our sample of stars for Sextans, finding a gradient in GSR of $4.1^{+2.6}_{-2.6}\kms\mathrm{deg}^{-1}$ at PA = $357^{+32}_{-32} \mathrm{deg}$. The gradient that we obtain is compatible with the one reported by \citet{Battaglia2011} at PA = 11 deg, and both of them are consistent with the direction of the PM in GSR of Sextans (PA = $333 \pm 2$ deg), which suggests that they could be due to the projection of the PM along the line of sight. As for the gradient reported along the semimajor axis of Sextans (\citealt{Battaglia2011}), we do not find any evidence, which could be explained by the different methodologies among the studies.
\end{itemize}

We also studied the velocity gradients due to the rotation in the plane of the sky of these galaxies. Their amplitudes, $A_{\mathrm{grad}}^{V_T}$, can be found in Table~\ref{tab:resultadosv}. We find significant clockwise rotation in Carina and Fornax. On the other hand, \citet{MartinezGarcia2021} reported significant clockwise rotation for Carina, Fornax, and also for Sculptor. The discrepancy with Sculptor is likely to be due to the different data sets used in both studies. In this study we use a data set with 3D velocities, which contains a lower number of sources and a more restricted spatial coverage. However, we should point out that the preferred direction of rotation found in this work for Sculptor is also clockwise (at $0.45\sigma$), which indicates that Sculptor appears to be rotating in clockwise sense in the plane of the sky.

\subsection{Environmental effects}

\begin{figure}
    \centering
    \includegraphics[width=\columnwidth]{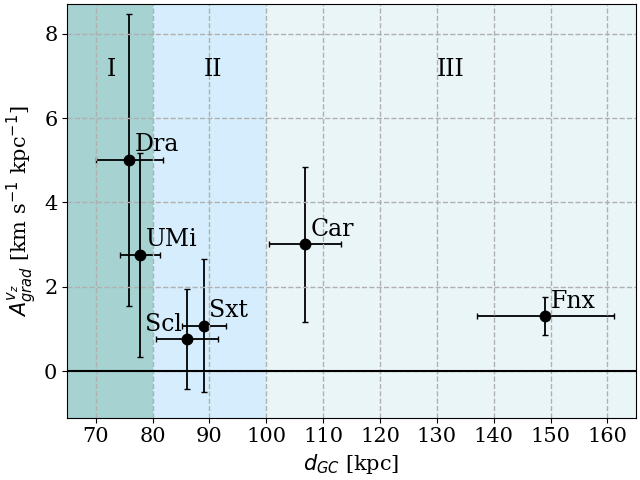}
    \caption{Comparison between the amplitude of the gradients in $v_z$ and the Galactocentric distance. Shaded areas represent galaxies with different behaviours; area marked as "I" is for galaxies that are currently moving towards their pericentre and whose gradients could to be induced by the MW, "II" for galaxies which have recently abandoned their pericentre and show no gradients since their internal kinematics are likely to be severely disrupted by the pericentric passage, and "III" for galaxies located at $\geq 100$ kpc.}
    \label{fig:d_GC_grad_los}
\end{figure}

The dwarfs analysed in this paper are expected to have interacted with the MW. This makes the  interpretation of the velocity gradients that we find in Carina, Draco, Fornax, and Ursa Minor not straightforward. They could be caused by the internal rotation of the galaxies or by other phenomena such as tidal effects. Such interaction could have left some traces in their internal kinematics. On the other hand, according to the tidal stirring scenario (\citealt{Mayer2010}), rotation-supported dwarfs would be repeatedly tidally shocked during successive pericentre passages and would be progressively transformed into pressure-supported systems. Should this scenario be correct, present day dSphs could show a remnant of their former rotation. However, this scenario seems to be in contradiction with the lack of evidence for rotation in the isolated dSphs  Cetus and Tucana (\citealt{Taibi2018, Taibi2020}). Both galaxies have old stellar populations (\citealt{Monelli2010a, Monelli2010b}), little gas (\citealt{Putman2021}), and at most have experienced a single pericentric passage around M31 or the MW. Therefore, these galaxies should have preserved a significant part of their original rotation if they were formerly rotating discs. Another point of friction is the fact that  the rotation versus dispersion support ($v_{rot} / \sigma$), does not show  significant differences between  satellites and isolated dwarfs of the LG (\citealt{Wheeler2017}).

\begin{figure}
    \centering
    \includegraphics[width=\linewidth]{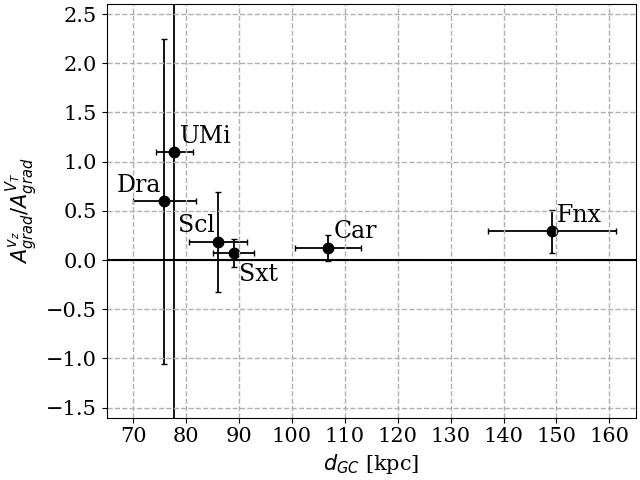}
    \caption{Comparison of the fraction of the amplitude of the gradients in $v_z$ and the amplitude due to rotation in the plane of the sky, $A_{\mathrm{grad}}^{v_z}/A_{\mathrm{grad}}^{V_T}$, and their Galactocentric distances.}
    \label{fig:d_GC_frac_gradient}
\end{figure}

Initially we explored possible relations between the gradients in $v_z$ and the distance of the satellites to the MW. We derive the Galactocentric distances assuming the Galactic centre to be located at $\alpha = 266.4051$ deg, $\delta = $--28.936175 deg (\citealt{Reid2004}), at distance of 8.3 kpc (\citealt{Gillessen2009}), and height of the Sun above the Galactic mid-plane of 27 pc (\citealt{Chen2001}). We use a MC scheme with $10^4$ iterations to determine the uncertainties of the distances. Galactocentric distances can be found in Table~\ref{tab:resultadosv}.
In Figure~\ref{fig:d_GC_grad_los}, we show a comparison of the amplitude of the gradients in $v_z$ and the Galactocentric distances. At first sight, there are no obvious trends except for some weak decreasing tendency.
The closest galaxies to the MW (i.e. Draco and Ursa Minor) show the largest non-negligible gradients. Carina and Fornax are located at distances $\geq 100$ kpc from the MW and also show non-negligible gradients, although their amplitude is lower. Sculptor and Sextans are located at intermediate Galactocentric distances, not much larger than those of Draco or Ursa Minor, and are not experiencing velocity gradients.  If the amplitude of the gradients was due exclusively to the present Galactocentric distance of the satellites, we would expect to find a progressive decrease of $A_{\mathrm{grad}}^{v_z}$ as the Galactocentric distance increases. However, given the weak trend and the change in the tendency that we observe for Sculptor and Sextans the amplitude of the gradients in $v_z$ does not seem to be exclusively determined by the current Galactocentric distance. In Figure~\ref{fig:d_GC_frac_gradient} we show a comparison of the fraction of $A_{\mathrm{grad}}^{v_z}/A_{\mathrm{grad}}^{V_T}$, and the Galactocentric distances of the galaxies. In this case, the galaxies from our sample seem to show a smoother trend in which systems closer to the MW show larger fractions of gradients in $v_z$ than in $v_T$. However, the large uncertainties measured in $A_{\mathrm{grad}}^{V_T}$ do not allow us to draw any strong conclusion. 

\begin{figure*}
    \centering
    \includegraphics[width=\linewidth
    ]{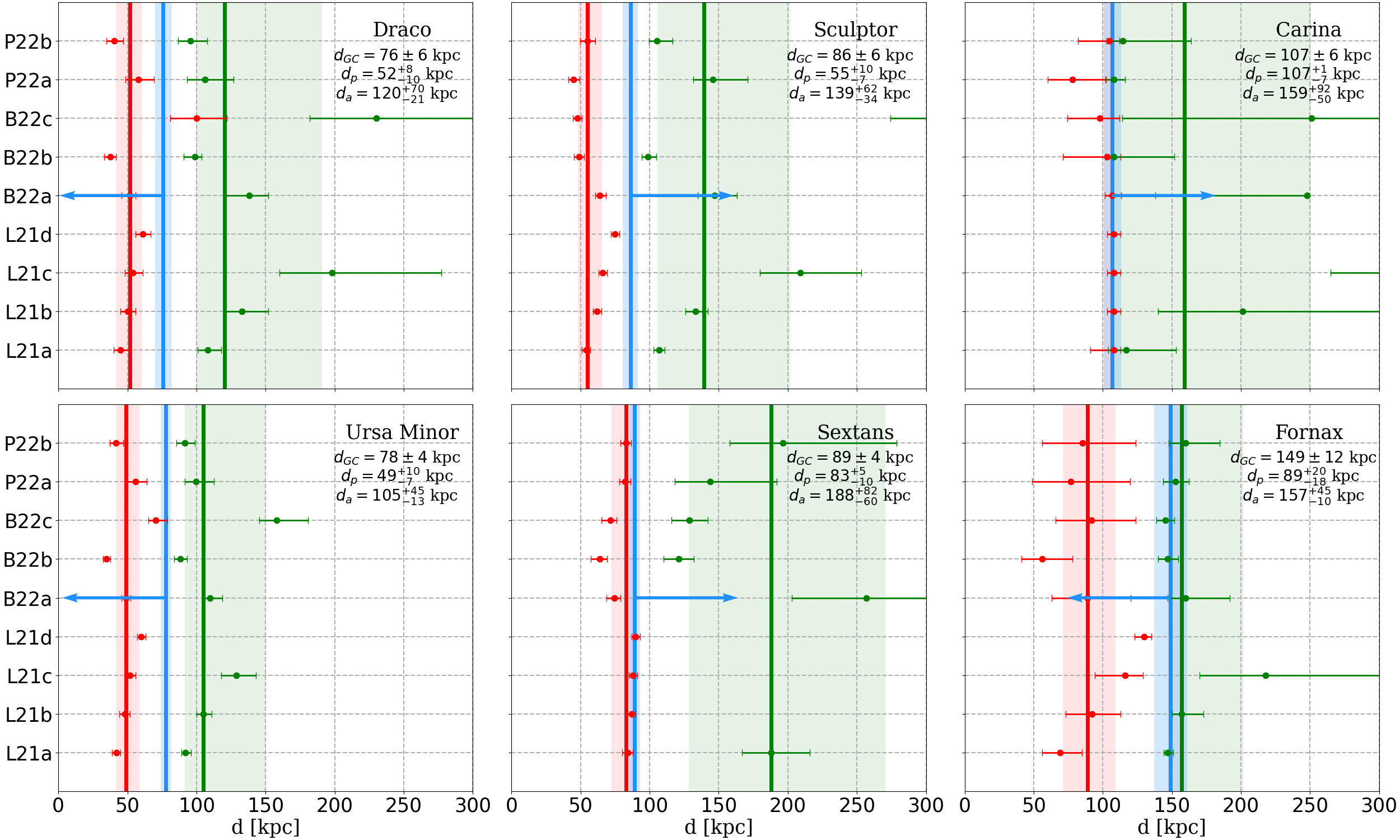}
    \caption{Orbital parameters taken from \citet{Li2021, Battaglia2022, Pace2022} and Galactocentric distances for the dSphs. Pericentric ($d_p$) and apocentric distances ($d_a$) are represented by red and green points, respectively. Their median values and the interval between the 0.16 and 0.84 quantiles are represented by vertical lines with shaded areas, with colours matching the corresponding parameters.  The current Galactocentric distance of the dwarfs are represented by the blue vertical line and shaded area. Blue horizontal arrows departing from the vertical line of the Galactocentric distance represent the direction of the radial velocity with respect to the MW. Arrows pointing to the left represent galaxies moving towards their pericentre, arrows pointing to the right represent galaxies moving towards their apocentres.  Markers in the y axis indicate the study from which the parameters have been taken:
    L21a, L21b, L21c, and L21d are for the PE$_{\mathrm{HM}}$, PNFW, PE$_{\mathrm{IM}}$, and PE$_{\mathrm{LM}}$ models from \citet{Li2021}; B22a, B22b, and B22c for Light MW, Heavy MW, and the perturbed potential models from \citet{Battaglia2022}; and P22a, and P22b for the MW+LMC, and MW models from \citet{Pace2022}, respectively.}
    \label{fig:orbits}
\end{figure*}

The current distance of the satellites to the MW is just a first approach to understand environmental effects. 
Orbital histories provide a  more accurate picture. In particular, information about the pericentre passages are is more informative, since the approach of a satellite to the pericentre is when the interaction with the host is more violent and can thus perturbate the internal kinematics of the system  (\citealt{Mayer2010}). In order to know whether these galaxies are approaching or getting away from their pericentres, we derive their radial velocities with respect to the centre of the MW. These Galactocentric radial velocities ($v_R^{GC}$) can be found in Table~\ref{tab:resultadosv}. A galaxy with positive $v_R^{GC}$ is moving away from the centre of the MW (i.e. it is moving towards the apocentre), on the contrary, a negative value implies that the galaxy is approaching the MW (i.e. it is moving towards the pericentre). 
We compare our values of $v_R^{GC}$ with the ones of \citet{Li2021}, finding that they are similar, with a mean difference of $\sim 4 \kms$ between studies.
We then combine the information of the Galactocentric distances and the velocities with respect to the MW with the reported orbital parameters for these dSphs that have been derived in \citet{Li2021, Battaglia2022}, and \cite{Pace2022} using different prescriptions and potentials. In Figure~\ref{fig:orbits} we show the different reported pericentric $(d_p)$ and apocentric $(d_a)$ distances for the dSphs, their current Galactocentric distances, and whether they are moving towards or away from their orbit's pericentre. We also show the median for $d_p$, and $d_a$ and the quantiles 0.16 and 0.84 of their distribution for the 9 analysed models.

With this information, it is possible to reassess Figure~\ref{fig:d_GC_grad_los}. First, for Draco and Ursa Minor (area marked as "I"), we find the largest non-negligible gradients in $v_z$.
Both galaxies are relatively close to the MW ($d_{GC} \sim 75$ kpc), and have similar pericentric distances ($\sim50$ kpc). This, combined with their negative Galactocentric radial velocities, $v_R^{GC} < 0 \kms$, indicates that they are relatively close to their pericentres and moving towards them. Given their proximity to the MW and the fact that they are approaching their pericentres, it is plausible that tidal forces exerted by the MW are causing the observed gradients in $v_z$. Since the Sun's position is relatively close to the MW's centre compared to the distances to the galaxies in our sample, the induced velocity gradient would be mostly projected along the line of sight, and thus detected as a gradient in $v_z$.
This scenario is consistent with Draco and Ursa Minor showing the largest values of $A_{\mathrm{grad}}^{v_z} / A_{\mathrm{grad}}^{V_T}$ (see Figure~\ref{fig:d_GC_frac_gradient}).
Secondly, the amplitude of the gradients in $v_z$ for Sculptor and Sextans (area marked as "II") is small, and compatible with zero within $1\sigma$.  These galaxies are currently moving away from their orbits pericentres ($v_R^{GC} > 0 \kms$). Such recent interaction with the MW is likely to have severely affected their internal kinematics. The tidal shock would have disrupted the kinematics of the systems, which would explain the absence of coherent kinematics among their stars.
Finally, Carina and Fornax (area marked as "III") are the furthermost galaxies, located at $\geq 100$ kpc of the MW, and both of them show non-negligible gradients in $v_z$. 

Carina's positive Galactocentric radial velocity, $v_R^{GC} = 8.79 \pm0.67\kms$, implies that the galaxy is slowly moving away from its pericentre. Moreover, the median pericentric distance of Carina inferred from the 9 orbit simulations ($d_p = 107^{+1}_{-7}$ kpc) is compatible with its current Galactocentric distance $(d_{GC} =106.8\pm6.3 \kpc)$. Given the large pericentric distance of its orbit (the largest in our sample), we expect tidal forces to have a moderate impact on Carina's internal kinematics following a pericentric passage. Furthermore, the resonant nature of some host satellite encounters (\citealt{Lokas2014}) could result in a stronger rotation signature in the dwarf right after the pericentre passage. Therefore, should Carina be in the vicinity of its orbit pericentre, the observed gradient could be due to the interaction with the MW. Nevertheless, we find that for four out of the nine orbital integrations that we analyse (in particular two of \citealt{Pace2022}, one of \citealt{Battaglia2022} and one of \citealt{Li2021}), the pericentric and apocentric distances are compatible among them and also with the Galactocentric distance (see Figure~\ref{fig:orbits}). The situation of Carina in its orbit is thus unclear. If Carina is in an orbit in which the pericentre and apocentre are close, the interaction with the MW would be unlikely to have caused the gradient due to the low ellipticity of the orbit. Either way, the large variance of the orbital parameters for Carina makes difficult to interpret the origin of its velocity gradient. Further studies on orbital integration may help to shed light on the possible impact of the MW on Carina's internal kinematics. In the case of Fornax, its internal gradient appears be due to either internal processes; i.e. internal rotation; or to a past interaction with the MW rather than to tidal forces currently affecting it. Fornax is the most distant galaxy in our sample $(d_{GC} = 149.1 \pm 12.1 \kpc)$ and is currently leaving its orbit apocentre ($d_{a} = 157^{+45}_{-10} \kpc$, $v_R^{GC} = -35.11 \pm0.31\kms$).

\begin{figure}
    \centering
    \includegraphics[width=1\linewidth]{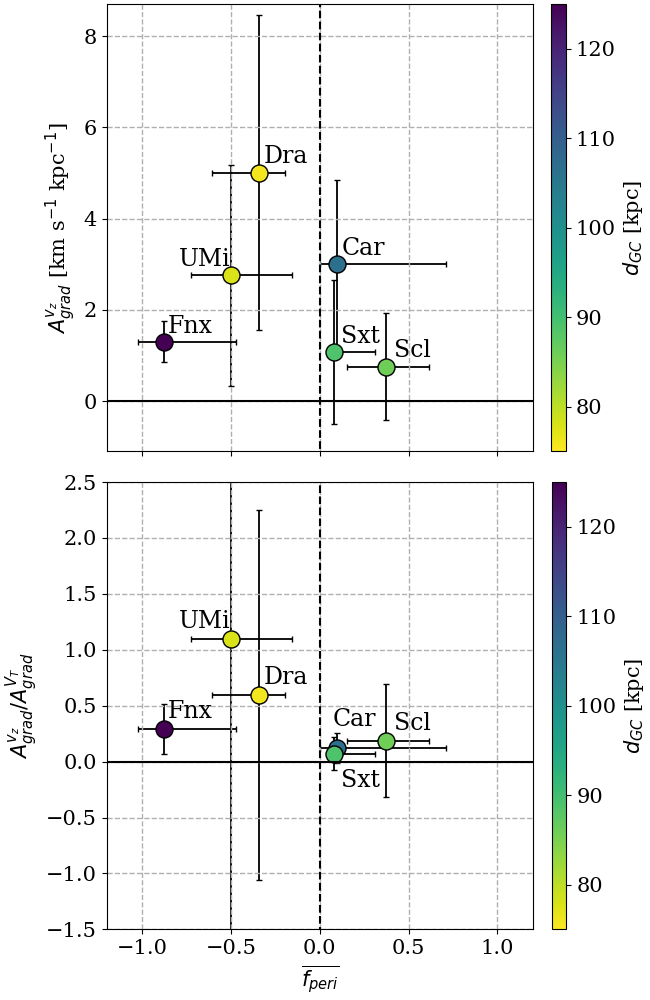}
    \caption{{\it Top}: Comparison between the amplitude of the gradients in $v_z$, $A^{v_z}_{\mathrm{grad}}$, and the relative position of the galaxy in its orbit, $\overline{f_{\mathrm{peri}}}$ (see text). The error bars in the $y$ axis represent the uncertainties in $A^{v_z}_{\mathrm{grad}}$. The error bars in the $x$ axis show the 0.16 and the 0.84 quantile of the distribution of $f_{\mathrm{peri}}$ for each galaxy using the different orbital models presented in Figure~\ref{fig:orbits}. The Galactocentric distance of the galaxies is marked in the colour bar.
    {\it Bottom}: Same plot but for the fraction between the gradient in $v_z$ and $v_T$, $A_{\mathrm{grad}}^{v_z} / A_{\mathrm{grad}}^{V_T}$.}
    \label{fig:frac}
\end{figure}

In Figure~\ref{fig:frac}, we show a comparison between the amplitude of the gradients in $v_z$ and $v_T$, and the relative distance of the galaxies with respect to their pericentres ($\overline{f_{\mathrm{peri}}}$), which summarizes well the aforementioned findings. We derive $\overline{f_{\mathrm{peri}}}$ as the median value of $f_{\mathrm{peri}} = \mathrm{sgn}(v_R^{GC}) (d_{GC} - d_p) / (d_a - d_p)$, where $d_{GC}$ is the Galactocentric distance of the galaxy. The orbits pericentric and apocentric distances, $d_p$ and $d_a$, are taken from the 9 different orbital models. The error bars show the 0.16 and 0.84 quantiles of the distribution of $f_{\mathrm{peri}}$ for each galaxy. The absolute value of $\overline{f_{\mathrm{peri}}}$ ranges between 0 (for a galaxy located in its pericentre) and 1 (for a galaxy located in its apocentre). The sign indicates whether the galaxy is approaching its orbit pericentre ($\overline{f_{\mathrm{peri}}} < 0$) or moving away from it ($\overline{f_{\mathrm{peri}}} > 0$). As $\overline{f_{\mathrm{peri}}}$ increases from negative values, galaxies tend to show larger values of $A_{\mathrm{grad}}^{v_z}$. These are galaxies that are approaching their orbit pericentres. Galaxies that have recently suffered a pericentre passage $\overline{f_{\mathrm{peri}}} > 0$ show, on average, smaller velocity gradients. As commented above, Carina seems to be an outlier. The galaxy could just have passed through its orbit pericentre and, given its relatively large distance to the MW, suffered mild tidal forces that caused the observed velocity gradient. Carina's projected optical major axis (PA = 65 deg) is aligned with its systemic motion (PA $\sim 67$ deg; \citealt{MartinezGarcia2021}). This together with the fact that the gradient we observe also shows a similar orientation (PA = $81^{+49}_{-49}$ deg), seem to support this scenario. However, the velocity gradient could also be produced by Carina's internal kinematics; i.e. internal rotation. Indeed Carina also shows a relatively strong rotation signal in the plane of the sky. This can be observed in the lower panel of Figure~\ref{fig:frac} where $A_{\mathrm{grad}}^{v_z} / A_{\mathrm{grad}}^{V_T} \sim 0.1$. Given the large dispersion we observe between different orbital models for Carina, and its relatively large pericentric distance, we do not make any strong claim about the origin of its velocity gradient.

\subsection{Environmental effects in TNG50}
We explored numerical simulations to assess our findings regarding the effect of the MW in the internal kinematics of dSphs. 
In short, we used the TNG50 run of  IllustrisTNG Project (\citealt{Marinacci2018, Naiman2018, Nelson2018, Pillepich2018b, Springel2018}), a suite of gravo-magnetohydrodynamic simulations which was run using the moving mesh \texttt{AREPO}  (\citealt{Springel2010}) code. The full description of the model for galaxy formation and evolution can be found in \citet{Weinberger2017, Pillepich2018a}. The adopted cosmology parameters are consistent with \citet{PlanckCollaboration2016}. Haloes are identified using friends-of-friends (FoF; \citealt{Davis1985}) algorithm and subhaloes using \texttt{SUBFIND} (\citealt{Springel2001, Dolag2009}). The simulation traces the evolution of the dark matter and stellar particles, stellar winds, gas cells, and black holes from $z = 127$ to $z = 0$, storing the data between $z\sim 20$ to $z = 0$ in 100 snapshots.
We used the TNG50-1 simulation (henceforth TNG50, \citealt{Nelson2019b, Pillepich2019}), the highest resolution run of the IllustrisTNG project. 
TNG50 is the best-suited simulation for our purposes, due to sizeable cosmological volume ($\sim 50$ Mpc co-moving box) and its resolution for baryons ($8.5 \times 10^4 M_{\odot}$ for gas cells and stellar particles). This allows studying the internal kinematics of a large sample of dwarf galaxies in the range of mass of interest.

\begin{figure*}
    \centering
    \includegraphics[width=\linewidth]{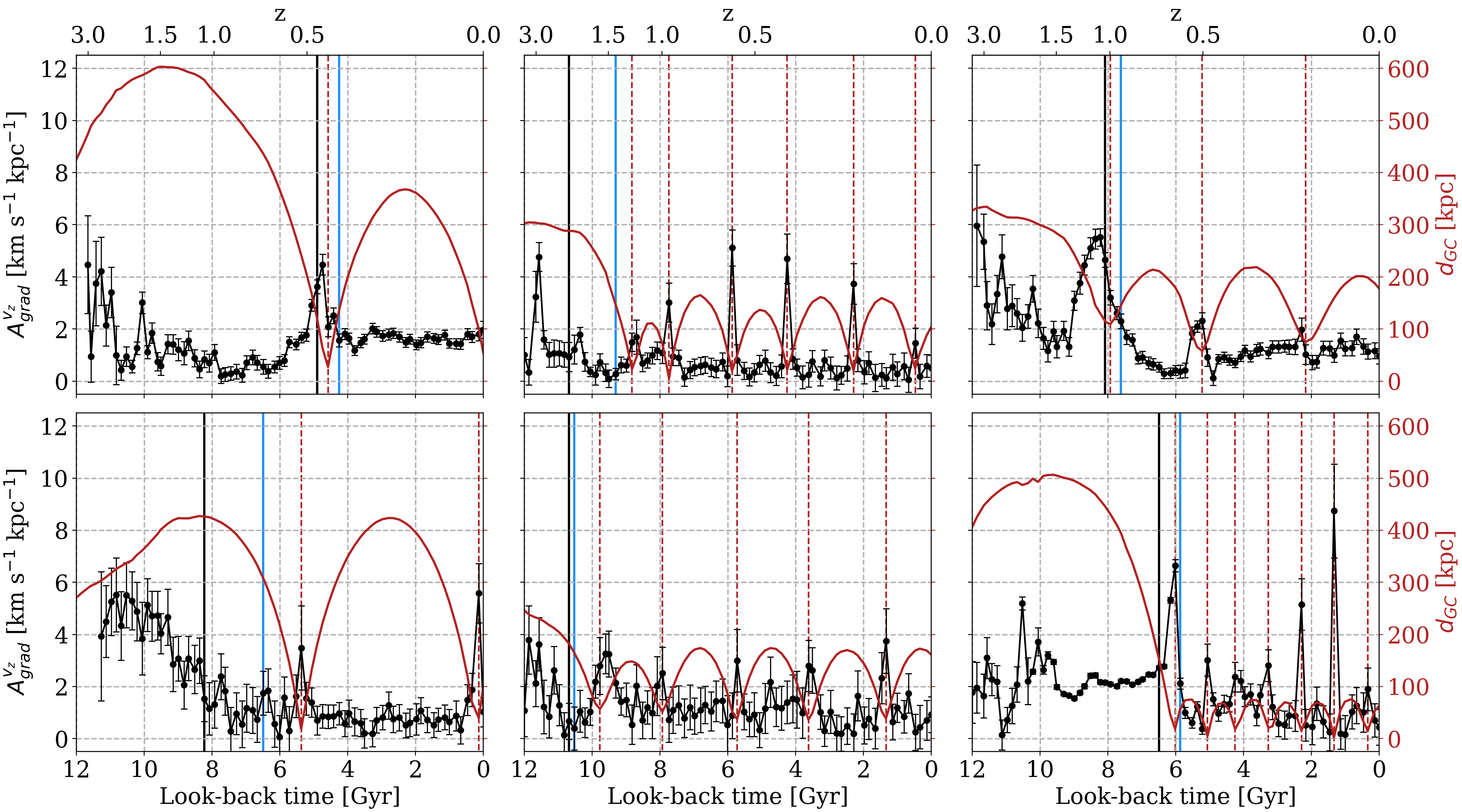}
    \caption{Evolution of $A_{\mathrm{grad}}^{v_z}$ across time for six dwarf satellites of TNG50 with similar orbital periods to the observed galaxies and pericentric distances ranging from $\sim5$ to $\sim110$ kpc. Black lines with error bars show the evolution of $A_{\mathrm{grad}}^{v_z}$ at different look-back times/redshifts. Red solid lines represent the distance of the dwarf galaxy to the host, with the pericentric passages marked as  dashed vertical lines. Blue vertical  lines represent the moment when a galaxy reached the 90 per cent of its cumulative star formation and black vertical lines the last merger suffered by the dwarf. There is an evident impact on the internal kinematics of the galaxies in each pericentric passage. These passages may lead to the disruption of the kinematics or slight increases of the gradients which are likely to be due to resonances (\citealt{Lokas2015}). Regardless of their initial kinematics, all the galaxies show a similar behaviour on the velocity gradients, with significant increases during the pericentric passages followed by drastic drops.}
    \label{fig:tng50}
\end{figure*}

We searched galaxies resembling the dwarfs of the LG. First, we looked for host galaxies similar to MW/M31 by selecting central subhaloes whose halo virial mass is in the range $0.7-3.0\times 10^{12} M_{\odot}$ at $z = 0$ (see \citealt{Patel2017a} and references therein). Thereupon, we filter their satellites imposing several criteria; the maximum circular velocity of the subhalo has to be $\leq 45 \kms$, the satellite has to be within the virial radius of the host at $z = 0$ (\citealt{Patel2018}), and the stellar mass in the range $M_{*} = 10^{7-10}M_{\odot}$ (this cut ensures $\geqslant 120$ stellar particles per galaxy, \citealt{Joshi2021}). We then exclude galaxies detected by \texttt{SUBFIND} which are not from cosmological origin and those which could not be traced up to at least $z = 2$ (\citealt{Joshi2021}). For the dwarf candidates, we derive their internal kinematics and gradients in $v_z$ proceeding analogously to Sections~\ref{subsec:internal_kinematics} and~\ref{sec:gradients}. 
We show six examples of galaxies and the evolution of their 
gradients in $v_z$ across time in Figure~\ref{fig:tng50}.
We show simulated dwarfs whose orbital periods are consistent with the ones derived by \citet{Battaglia2022} for the galaxies of our sample (see Table~\ref{tab:resultadosv}). In particular we show dwarfs with a few pericente passages (left-hand panels), 
and dwarfs which have performed multiple pericentric passages, with pericentric distances ranging from to $\sim5$ to $\sim110$ kpc (central and right-hand panels). In each panel we show when the galaxies reached the 90 per cent of their cumulative  stellar formation history (blue line) and the moment of their last merger (black line).
We observe a clear increase in $A_{\mathrm{grad}}^{v_z}$ when the galaxies approach the pericentre of their trajectories, followed by a drastic reduction of $A_{\mathrm{grad}}^{v_z}$ as they get away from it. The increase seems to be more progressive in the first pericentres, followed by sharper increases as the galaxies perform more passages. Additionally, we observe that the more pericentric passages a galaxy experiences, and the closer to the host, the lower $A_{\mathrm{grad}}^{v_z}$ is when the galaxy abandons the pericentres. This can be related to the original internal kinematics of the galaxy; the first pericentres would disrupt significantly their kinematic patterns, and subsequent passages would remove them almost completely. Thus, the random motions of the final perturbed system would be more affected very close to the pericentre.

Besides the interaction with the host, other mechanisms could be invoked to try to explain these increases of the velocity gradients, such as mergers or stellar formation. The selected dwarfs from TNG50 have stellar masses significantly larger than those of the observed satellites, given the limited resolution of the simulation and due to the fact that we require at least 100 stellar particles to study the internal kinematics. These larger stellar masses could imply stellar formation histories (SFH) different from the ones of the observed dSphs. However, the analysis of the SFH of the simulated dwarfs shows that they reach the 90 per cent of their cumulative SFH, on average, at $z \sim 1.0 \pm 0.5$, being consistent with the SFH of the dSphs of the LG (\citealt{Weisz2014}). The 90 per cent of the SFH is usually reached before or right after the first infall of the galaxy. As for the mergers, we have analysed the merger trees provided by TNG50 for the simulated dwarfs, finding that the vast majority of them happen before the galaxy reached the pericentre in the first infall of the galaxy to the host. 
Therefore, mergers or the stellar formation do not seem to be the main drivers of the increases of the present day velocity gradients.
On the other hand, we observe that irrespective of the original kinematics of the dwarfs (some of them with high and others with low velocity gradients), the successive pericentre passages lead to similar behaviours of the velocity gradients, with significant increases during the pericentric passages followed by sharp decreases as the dwarfs abandon the pericentre. This suggests that the interaction with the host is the main driver of the present day velocity gradients 
The influence of the host galaxy on the internal kinematics of its satellites observed in TNG50 is in agreement with our observations in the MW dwarfs, i.e. Draco and Ursa Minor showing non-negligible gradients as they head towards their pericentres and no gradients in Sculptor and Sextans as they leave them.

\section{Conclusions}
\label{sec:conclusions}

In this paper, we have studied the internal kinematics of six MW dSph satellites, namely, Carina, Draco, Fornax, Sculptor, Sextans, and Ursa Minor. We selected member stars from \textit{Gaia} DR3 and combined them with $\vlos$ available in the literature. We derived the 3D internal kinematics of their stars and studied the presence of internal velocity gradients correcting for perspective effects. To do so, we introduced a co-moving orthogonal reference frame centred on each galaxy CM, with the $z$ axis pointing along the line-of-sight at the centre of the galaxy. We found non-negligible gradients in $v_z$ ($A_{\rm grad}^{v_z} > 0$) with $\geq 1\sigma$ level of significance for Carina, Draco, Fornax, and Ursa Minor. We did not find velocity gradients in $v_z$ for Sculptor and Sextans, what suggests that these two galaxies do not rotate along the line-of-sight.

We studied possible interactions of these dSphs with the MW and how they may have impacted their internal kinematics. We found that galaxies whose orbits have small pericentre distances that are now approaching to such pericentre (Draco and Ursa Minor) show larger velocity gradients along the line-of-sight than those leaving the pericentre (Sculptor and, Sextans), or those located at larger distances (Carina and Fornax). In fact, the galaxies leaving their pericentres showed the lowest velocity gradient in $v_z$, followed by those located at large distances. We compared such gradients with the rotation observed in the plane of the sky for the same sample of stars, $A_{\rm grad}^{V_T}$, and found that in general, satellites closer to the MW tend to show larger $A_{\rm grad}^{v_z}/ A_{\rm grad}^{V_T}$ ratios.

We assessed our findings analysing dwarf satellites of MW/M31-like hosts in the TNG50 simulation. We found that the simulated dwarfs show similar behaviours to the observed ones, with a clear increase of their velocity gradients in $v_z$ when they approach their pericentres, followed by a drastic decrease as the galaxies leave them. This behaviour is common to all the galaxies we analysed in TNG50, regardless of their initial kinematic state, star formation histories or merger histories, which suggests that their internal kinematics is most influenced by the gravitational interaction with the host galaxy. We therefore conclude that the line-of-sight gradients currently observed in the MW satellites are largely caused by the MW gravitational force. During their approach to the MW, tidal forces would induce a torque that creates a velocity gradient within the body of the dwarfs. This could be the case of Draco and Ursa Minor, which are now approaching their pericentres. The intense forces suffered by the dwarfs during their pericentre would stir their content, removing most of the coherent motion of their stars. This may be the case of Sculptor and Sextans, which just passed through their pericentres. For galaxies located at larger distances, such Carina or Fornax, it seems unlikely that the observed gradient in $v_z$ is caused by the current interaction with the MW, but rather a past interaction with the MW or other systems, or the result of the internal rotation of these systems.

\section*{Acknowledgements}
The authors thank the anonymous referee for the comments that have helped to improve this paper.
We acknowledge support from the Spanish Agencia Estatal de Investigación del Ministerio de Ciencia e Innovacion (AEI-MICINN) under grant "Proyectos de I+D+i" with references AYA2017-89841-P and PID2020-115981GB-I00. AdP acknowledges the financial support from the European Union - NextGenerationEU and the Spanish Ministry of Science and Innovation through the Recovery and Resilience Facility project J-CAVA. The authors also acknowledge all the open-source software involved in this study, specially TOPCAT, Python, PostgreSQL and git. This work has made use of data from the European Space Agency (ESA) mission {\it Gaia} (\url{https://www.cosmos.esa.int/gaia}), processed by the {\it Gaia} Data Processing and Analysis Consortium (DPAC; \url{https://www.cosmos.esa.int/web/gaia/dpac/consortium}). Funding for the DPAC has been provided by national institutions, in particular, the institutions participating in the {\it Gaia} Multilateral Agreement.
 
\section*{Data availability}
All the data underlying this article are publicly available. The \textit{Gaia} DR3 data can  be extracted from the \textit{Gaia} archive (\url{https://gea.esac.esa.int/archive}). The software for the membership selection is available at GitHub (\url{https://github.com/AndresdPM/GetGaia}). 
The radial velocity catalogues were taken from \citet{Walker2009, Walker2015, Pace2020, Pace2021}. The IllustrisTNG simulations are publicly available and can be found in \url{www.tng-project.org/data}. 
Catalogues including radial and tangential velocities for member stars will be shared on reasonable request to the corresponding author.



\bibliographystyle{mnras}
\bibliography{biblio.bib} 

\begin{thebibliography}{}
\makeatletter
\relax
\def\mn@urlcharsother{\let\do\@makeother \do\$\do\&\do\#\do\^\do\_\do\%\do\~}
\def\mn@doi{\begingroup\mn@urlcharsother \@ifnextchar [ {\mn@doi@}
  {\mn@doi@[]}}
\def\mn@doi@[#1]#2{\def\@tempa{#1}\ifx\@tempa\@empty \href
  {http://dx.doi.org/#2} {doi:#2}\else \href {http://dx.doi.org/#2} {#1}\fi
  \endgroup}
\def\mn@eprint#1#2{\mn@eprint@#1:#2::\@nil}
\def\mn@eprint@arXiv#1{\href {http://arxiv.org/abs/#1} {{\tt arXiv:#1}}}
\def\mn@eprint@dblp#1{\href {http://dblp.uni-trier.de/rec/bibtex/#1.xml}
  {dblp:#1}}
\def\mn@eprint@#1:#2:#3:#4\@nil{\def\@tempa {#1}\def\@tempb {#2}\def\@tempc
  {#3}\ifx \@tempc \@empty \let \@tempc \@tempb \let \@tempb \@tempa \fi \ifx
  \@tempb \@empty \def\@tempb {arXiv}\fi \@ifundefined
  {mn@eprint@\@tempb}{\@tempb:\@tempc}{\expandafter \expandafter \csname
  mn@eprint@\@tempb\endcsname \expandafter{\@tempc}}}

\bibitem[\protect\citeauthoryear{Amorisco \& Evans}{Amorisco \&
  Evans}{2012}]{Amorisco2012}
Amorisco N.~C.,  Evans N.~W.,  2012, \mn@doi [\mnras]
  {10.1111/j.1365-2966.2011.19684.x}, 419, 184

\bibitem[\protect\citeauthoryear{Aparicio, Carrera  \&
  Mart{\'{i}}nez-Delgado}{Aparicio et~al.}{2001}]{Aparicio2001}
Aparicio A.,  Carrera R.,   Mart{\'{i}}nez-Delgado D.,  2001, \mn@doi [\aj]
  {10.1086/323535}, 122, 2524

\bibitem[\protect\citeauthoryear{{Battaglia} \& {Nipoti}}{{Battaglia} \&
  {Nipoti}}{2022}]{Battaglia2022b}
{Battaglia} G.,  {Nipoti} C.,  2022, arXiv e-prints, \href
  {https://ui.adsabs.harvard.edu/abs/2022arXiv220507821B} {p. arXiv:2205.07821}

\bibitem[\protect\citeauthoryear{Battaglia et~al.,}{Battaglia
  et~al.}{2006}]{Battaglia2006}
Battaglia G.,  et~al., 2006, \mn@doi [\aap] {10.1051/0004-6361:20065720}, 459,
  423

\bibitem[\protect\citeauthoryear{Battaglia, Irwin, Tolstoy, Hill, Helmi,
  Letarte  \& Jablonka}{Battaglia et~al.}{2008a}]{Battaglia2008a}
Battaglia G.,  Irwin M.,  Tolstoy E.,  Hill V.,  Helmi A.,  Letarte B.,
  Jablonka P.,  2008a, \mn@doi [\mnras] {10.1111/j.1365-2966.2007.12532.x},
  383, 183

\bibitem[\protect\citeauthoryear{Battaglia, Helmi, Tolstoy, Irwin, Hill  \&
  Jablonka}{Battaglia et~al.}{2008b}]{Battaglia2008b}
Battaglia G.,  Helmi A.,  Tolstoy E.,  Irwin M.,  Hill V.,   Jablonka P.,
  2008b, \mn@doi [\apj] {10.1086/590179}, 681, L13

\bibitem[\protect\citeauthoryear{Battaglia, Tolstoy, Helmi, Irwin, Parisi, Hill
   \& Jablonka}{Battaglia et~al.}{2011}]{Battaglia2011}
Battaglia G.,  Tolstoy E.,  Helmi A.,  Irwin M.,  Parisi P.,  Hill V.,
  Jablonka P.,  2011, \mn@doi [\mnras] {10.1111/j.1365-2966.2010.17745.x}, 411,
  1013

\bibitem[\protect\citeauthoryear{{Battaglia}, {Taibi}, {Thomas}  \&
  {Fritz}}{{Battaglia} et~al.}{2022}]{Battaglia2022}
{Battaglia} G.,  {Taibi} S.,  {Thomas} G.~F.,   {Fritz} T.~K.,  2022, \mn@doi
  [\aap] {10.1051/0004-6361/202141528}, \href
  {https://ui.adsabs.harvard.edu/abs/2022A&A...657A..54B} {657, A54}

\bibitem[\protect\citeauthoryear{Bettinelli, Hidalgo, Cassisi, Aparicio  \&
  Piotto}{Bettinelli et~al.}{2018}]{Bettinelli2018}
Bettinelli M.,  Hidalgo S.~L.,  Cassisi S.,  Aparicio A.,   Piotto G.,  2018,
  \mn@doi [\mnras] {10.1093/mnras/sty226}, 476, 71

\bibitem[\protect\citeauthoryear{Bettinelli, Hidalgo, Cassisi, Aparicio,
  Piotto, Valdes  \& Walker}{Bettinelli et~al.}{2019}]{Bettinelli2019}
Bettinelli M.,  Hidalgo S.~L.,  Cassisi S.,  Aparicio A.,  Piotto G.,  Valdes
  F.,   Walker A.~R.,  2019, \mn@doi [\mnras] {10.1093/mnras/stz1679}, 487,
  5862

\bibitem[\protect\citeauthoryear{Blumenthal, Faber, Primack  \&
  Rees}{Blumenthal et~al.}{1984}]{Blumenthal1984}
Blumenthal G.~R.,  Faber S.~M.,  Primack J.~R.,   Rees M.~J.,  1984, \mn@doi
  [Nature] {10.1038/311517a0}, 311, 517

\bibitem[\protect\citeauthoryear{Cappellari \& Copin}{Cappellari \&
  Copin}{2003}]{Capellari2003}
Cappellari M.,  Copin Y.,  2003, \mn@doi [\mnras]
  {10.1046/j.1365-8711.2003.06541.x}, 342, 345

\bibitem[\protect\citeauthoryear{Carrera, Aparicio, Mart{\'{i}}nez-Delgado  \&
  Alonso-Garc{\'{i}}a}{Carrera et~al.}{2002}]{Carrera2002}
Carrera R.,  Aparicio A.,  Mart{\'{i}}nez-Delgado D.,   Alonso-Garc{\'{i}}a J.,
   2002, \mn@doi [\aj] {10.1086/340702}, 123, 3199

\bibitem[\protect\citeauthoryear{Chen et~al.,}{Chen et~al.}{2001}]{Chen2001}
Chen B.,  et~al., 2001, \mn@doi [\apj] {10.1086/320647}, 553, 184

\bibitem[\protect\citeauthoryear{{Davis}, {Efstathiou}, {Frenk}  \&
  {White}}{{Davis} et~al.}{1985}]{Davis1985}
{Davis} M.,  {Efstathiou} G.,  {Frenk} C.~S.,   {White} S.~D.~M.,  1985,
  \mn@doi [\apj] {10.1086/163168}, \href
  {https://ui.adsabs.harvard.edu/abs/1985ApJ...292..371D} {292, 371}

\bibitem[\protect\citeauthoryear{Dekel \& Silk}{Dekel \&
  Silk}{1986}]{Dekel1986}
Dekel A.,  Silk J.,  1986, \mn@doi [ApJ] {10.1086/164050}, 303, 39

\bibitem[\protect\citeauthoryear{Dolag, Borgani, Murante  \& Springel}{Dolag
  et~al.}{2009}]{Dolag2009}
Dolag K.,  Borgani S.,  Murante G.,   Springel V.,  2009, \mn@doi [\mnras]
  {10.1111/j.1365-2966.2009.15034.x}, 399, 497

\bibitem[\protect\citeauthoryear{Fabricius et~al.,}{Fabricius
  et~al.}{2021}]{Fabricius2020}
Fabricius C.,  et~al., 2021, \mn@doi [A\&A] {10.1051/0004-6361/202039834}, 649,
  A5

\bibitem[\protect\citeauthoryear{Fabrizio et~al.,}{Fabrizio
  et~al.}{2016}]{Fabrizio2016}
Fabrizio M.,  et~al., 2016, \mn@doi [\apj] {10.3847/0004-637x/830/2/126}, 830,
  126

\bibitem[\protect\citeauthoryear{{Feast}, {Thackeray}  \& {Wesselink}}{{Feast}
  et~al.}{1961}]{Feast1961}
{Feast} M.~W.,  {Thackeray} A.~D.,   {Wesselink} A.~J.,  1961, \mn@doi [\mnras]
  {10.1093/mnras/122.5.433}, \href
  {https://ui.adsabs.harvard.edu/abs/1961MNRAS.122..433F} {122, 433}

\bibitem[\protect\citeauthoryear{{Gaia Collaboration} et~al.,}{{Gaia
  Collaboration} et~al.}{2016}]{GaiaCollaboration2016}
{Gaia Collaboration} et~al., 2016, \mn@doi [\aap]
  {10.1051/0004-6361/201629272}, 595, A1

\bibitem[\protect\citeauthoryear{{Gaia Collaboration} et~al.,}{{Gaia
  Collaboration} et~al.}{2021}]{GaiaEDR3}
{Gaia Collaboration} et~al., 2021, \mn@doi [A\&A]
  {10.1051/0004-6361/202039657}, 649, A1

\bibitem[\protect\citeauthoryear{{Gaia Collaboration}, {Vallenari, A.}, {Brown,
  A.G.A.}, {Prusti, T.}  \& {et al.}}{{Gaia Collaboration}
  et~al.}{2022}]{GaiaDR3}
{Gaia Collaboration} {Vallenari, A.} {Brown, A.G.A.} {Prusti, T.}  {et al.}
  2022, \mn@doi [A\&A] {10.1051/0004-6361/202243940}

\bibitem[\protect\citeauthoryear{{Gallagher} \& Wyse}{{Gallagher} \&
  Wyse}{1994}]{Gallagher1994}
{Gallagher} J. S. I. I.~I.,  Wyse R. F.~G.,  1994, \pasp, 106, 1225

\bibitem[\protect\citeauthoryear{Gillessen, Eisenhauer, Trippe, Alexander,
  Genzel, Martins  \& Ott}{Gillessen et~al.}{2009}]{Gillessen2009}
Gillessen S.,  Eisenhauer F.,  Trippe S.,  Alexander T.,  Genzel R.,  Martins
  F.,   Ott T.,  2009, \mn@doi [\apj] {10.1088/0004-637x/692/2/1075}, 692, 1075

\bibitem[\protect\citeauthoryear{Irwin \& Hatzidimitriou}{Irwin \&
  Hatzidimitriou}{1995}]{Irwin1995}
Irwin M.,  Hatzidimitriou D.,  1995, \mn@doi [\mnras]
  {10.1093/mnras/277.4.1354}, 277, 1354

\bibitem[\protect\citeauthoryear{Joshi, Pillepich, Nelson, Zinger, Marinacci,
  Springel, Vogelsberger  \& Hernquist}{Joshi et~al.}{2021}]{Joshi2021}
Joshi G.~D.,  Pillepich A.,  Nelson D.,  Zinger E.,  Marinacci F.,  Springel
  V.,  Vogelsberger M.,   Hernquist L.,  2021, \mn@doi [\mnras]
  {10.1093/mnras/stab2573}, 508, 1652

\bibitem[\protect\citeauthoryear{Kazantzidis, {\L}okas, Callegari, Mayer  \&
  Moustakas}{Kazantzidis et~al.}{2011}]{Kazantzidis2011}
Kazantzidis S.,  {\L}okas E.~L.,  Callegari S.,  Mayer L.,   Moustakas L.~A.,
  2011, \mn@doi [\apj] {10.1088/0004-637x/726/2/98}, 726, 98

\bibitem[\protect\citeauthoryear{Kirby et~al.,}{Kirby et~al.}{2010}]{Kirby2010}
Kirby E.~N.,  et~al., 2010, \mn@doi [\apjs] {10.1088/0067-0049/191/2/352}, 191,
  352

\bibitem[\protect\citeauthoryear{Kirby, Cohen, Guhathakurta, Cheng, Bullock  \&
  Gallazzi}{Kirby et~al.}{2013}]{Kirby2013}
Kirby E.~N.,  Cohen J.~G.,  Guhathakurta P.,  Cheng L.,  Bullock J.~S.,
  Gallazzi A.,  2013, \mn@doi [\apj] {10.1088/0004-637x/779/2/102}, 779, 102

\bibitem[\protect\citeauthoryear{Kirby et~al.,}{Kirby et~al.}{2015}]{Kirby2015}
Kirby E.~N.,  et~al., 2015, \mn@doi [\apj] {10.1088/0004-637x/801/2/125}, 801,
  125

\bibitem[\protect\citeauthoryear{Kleyna, Wilkinson, Evans, Gilmore  \&
  Frayn}{Kleyna et~al.}{2002}]{Kleyna2002}
Kleyna J.,  Wilkinson M.~I.,  Evans N.~W.,  Gilmore G.,   Frayn C.,  2002,
  \mn@doi [\mnras] {10.1046/j.1365-8711.2002.05155.x}, 330, 792

\bibitem[\protect\citeauthoryear{Koch, Grebel, Wyse, Kleyna, Wilkinson,
  Harbeck, Gilmore  \& Evans}{Koch et~al.}{2006a}]{Koch2006a}
Koch A.,  Grebel E.~K.,  Wyse R. F.~G.,  Kleyna J.~T.,  Wilkinson M.~I.,
  Harbeck D.~R.,  Gilmore G.~F.,   Evans N.~W.,  2006a, \mn@doi [\aj]
  {10.1086/499490}, 131, 895

\bibitem[\protect\citeauthoryear{Koch, Grebel, Kleyna, Wilkinson, Harbeck,
  Gilmore, Wyse  \& Evans}{Koch et~al.}{2006b}]{Koch2006b}
Koch A.,  Grebel E.~K.,  Kleyna J.~T.,  Wilkinson M.~I.,  Harbeck D.~R.,
  Gilmore G.~F.,  Wyse R. F.~G.,   Evans N.~W.,  2006b, \mn@doi [\aj]
  {10.1086/509889}, 133, 270

\bibitem[\protect\citeauthoryear{Koch, Kleyna, Wilkinson, Grebel, Gilmore,
  Evans, Wyse  \& Harbeck}{Koch et~al.}{2007a}]{Koch2007b}
Koch A.,  Kleyna J.~T.,  Wilkinson M.~I.,  Grebel E.~K.,  Gilmore G.~F.,  Evans
  N.~W.,  Wyse R. F.~G.,   Harbeck D.~R.,  2007a, \mn@doi [\aj]
  {10.1086/519380}, 134, 566

\bibitem[\protect\citeauthoryear{Koch, Wilkinson, Kleyna, Gilmore, Grebel,
  Mackey, Evans  \& Wyse}{Koch et~al.}{2007b}]{Koch2007a}
Koch A.,  Wilkinson M.~I.,  Kleyna J.~T.,  Gilmore G.~F.,  Grebel E.~K.,
  Mackey A.~D.,  Evans N.~W.,   Wyse R. F.~G.,  2007b, \mn@doi [\apj]
  {10.1086/510879}, 657, 241

\bibitem[\protect\citeauthoryear{Li, Hammer, Babusiaux, Pawlowski, Yang,
  Arenou, Du  \& Wang}{Li et~al.}{2021}]{Li2021}
Li H.,  Hammer F.,  Babusiaux C.,  Pawlowski M.~S.,  Yang Y.,  Arenou F.,  Du
  C.,   Wang J.,  2021, \mn@doi [\apj] {10.3847/1538-4357/ac0436}, 916, 8

\bibitem[\protect\citeauthoryear{{Lindegren} et~al.,}{{Lindegren}
  et~al.}{2021a}]{Lindegren2020}
{Lindegren} L.,  et~al., 2021a, \mn@doi [\aap] {10.1051/0004-6361/202039709},
  \href {https://ui.adsabs.harvard.edu/abs/2021A&A...649A...2L} {649, A2}

\bibitem[\protect\citeauthoryear{{Lindegren} et~al.,}{{Lindegren}
  et~al.}{2021b}]{Lindegren2020b}
{Lindegren} L.,  et~al., 2021b, \mn@doi [\aap] {10.1051/0004-6361/202039653},
  \href {https://ui.adsabs.harvard.edu/abs/2021A&A...649A...4L} {649, A4}

\bibitem[\protect\citeauthoryear{{{\L}okas}, {Athanassoula}, {Debattista},
  {Valluri}, {Pino}, {Semczuk}, {Gajda}  \& {Kowalczyk}}{{{\L}okas}
  et~al.}{2014}]{Lokas2014}
{{\L}okas} E.~L.,  {Athanassoula} E.,  {Debattista} V.~P.,  {Valluri} M.,
  {Pino} A.~d.,  {Semczuk} M.,  {Gajda} G.,   {Kowalczyk} K.,  2014, \mn@doi
  [\mnras] {10.1093/mnras/stu1846}, \href
  {https://ui.adsabs.harvard.edu/abs/2014MNRAS.445.1339L} {445, 1339}

\bibitem[\protect\citeauthoryear{{\L}okas, Semczuk, Gajda  \&
  D'Onghia}{{\L}okas et~al.}{2015}]{Lokas2015}
{\L}okas E.~L.,  Semczuk M.,  Gajda G.,   D'Onghia E.,  2015, \mn@doi [\apj]
  {10.1088/0004-637x/810/2/100}, 810, 100

\bibitem[\protect\citeauthoryear{Marinacci et~al.,}{Marinacci
  et~al.}{2018}]{Marinacci2018}
Marinacci F.,  et~al., 2018, \mn@doi [\mnras] {10.1093/mnras/sty2206}, 480,
  5113

\bibitem[\protect\citeauthoryear{Martínez-García, del Pino, Aparicio,
  van der Marel  \& Watkins}{Martínez-García
  et~al.}{2021}]{MartinezGarcia2021}
Martínez-García A.~M.,  del Pino A.,  Aparicio A.,  van der Marel R.~P.,
  Watkins L.~L.,  2021, \mn@doi [\mnras] {10.1093/mnras/stab1568}, 505, 5884

\bibitem[\protect\citeauthoryear{Massari, Breddels, Helmi, Posti, Brown  \&
  Tolstoy}{Massari et~al.}{2018}]{Massari2018}
Massari D.,  Breddels M.,  Helmi A.,  Posti L.,  Brown A.,   Tolstoy E.,  2018,
  Nature Astronomy, 2, 156

\bibitem[\protect\citeauthoryear{Massari, Helmi, Mucciarelli, Sales, Spina  \&
  Tolstoy}{Massari et~al.}{2020}]{Massari2020}
Massari D.,  Helmi A.,  Mucciarelli A.,  Sales L.~V.,  Spina L.,   Tolstoy E.,
  2020, \mn@doi [A\&A] {10.1051/0004-6361/201935613}, 633, A36

\bibitem[\protect\citeauthoryear{Mayer}{Mayer}{2010}]{Mayer2010}
Mayer L.,  2010, \mn@doi [Advances in Astronomy] {10.1155/2010/278434}, 2010,
  278434

\bibitem[\protect\citeauthoryear{McConnachie}{McConnachie}{2012}]{McConnachie2012}
McConnachie A.~W.,  2012, \mn@doi [\aj] {10.1088/0004-6256/144/1/4}, 144

\bibitem[\protect\citeauthoryear{McConnachie \& Venn}{McConnachie \&
  Venn}{2020a}]{McConnachie2020b}
McConnachie A.~W.,  Venn K.~A.,  2020a, Research Notes of the AAS, 4, 229

\bibitem[\protect\citeauthoryear{McConnachie \& Venn}{McConnachie \&
  Venn}{2020b}]{McConnachie2020a}
McConnachie A.~W.,  Venn K.~A.,  2020b, \aj, 160, 124

\bibitem[\protect\citeauthoryear{Monelli et~al.,}{Monelli
  et~al.}{2010a}]{Monelli2010a}
Monelli M.,  et~al., 2010a, \mn@doi [\apj] {10.1088/0004-637x/720/2/1225}, 720,
  1225

\bibitem[\protect\citeauthoryear{Monelli et~al.,}{Monelli
  et~al.}{2010b}]{Monelli2010b}
Monelli M.,  et~al., 2010b, \mn@doi [\apj] {10.1088/0004-637x/722/2/1864}, 722,
  1864

\bibitem[\protect\citeauthoryear{{Mu{\~n}oz} et~al.,}{{Mu{\~n}oz}
  et~al.}{2006}]{Munoz2006}
{Mu{\~n}oz} R.~R.,  et~al., 2006, \mn@doi [\apj] {10.1086/505620}, \href
  {https://ui.adsabs.harvard.edu/abs/2006ApJ...649..201M} {649, 201}

\bibitem[\protect\citeauthoryear{Mu{\~{n}}oz et~al.,}{Mu{\~{n}}oz
  et~al.}{2005}]{Munoz2005}
Mu{\~{n}}oz R.~R.,  et~al., 2005, \mn@doi [\apj] {10.1086/497396}, 631, L137

\bibitem[\protect\citeauthoryear{Naiman et~al.,}{Naiman
  et~al.}{2018}]{Naiman2018}
Naiman J.~P.,  et~al., 2018, \mn@doi [\mnras] {10.1093/mnras/sty618}, 477, 1206

\bibitem[\protect\citeauthoryear{Navarro, Frenk  \& White}{Navarro
  et~al.}{1995}]{NavarroFrenkWhite1995}
Navarro J.~F.,  Frenk C.~S.,   White S. D.~M.,  1995, \mn@doi [\mnras]
  {10.1093/mnras/275.3.720}, 275, 720

\bibitem[\protect\citeauthoryear{Nelson et~al.,}{Nelson
  et~al.}{2018}]{Nelson2018}
Nelson D.,  et~al., 2018, \mn@doi [\mnras] {10.1093/mnras/stx3040}, 475, 624

\bibitem[\protect\citeauthoryear{Nelson et~al.,}{Nelson
  et~al.}{2019}]{Nelson2019b}
Nelson D.,  et~al., 2019, \mn@doi [\mnras] {10.1093/mnras/stz2306}, 490, 3234

\bibitem[\protect\citeauthoryear{Pace et~al.,}{Pace et~al.}{2020}]{Pace2020}
Pace A.~B.,  et~al., 2020, \mn@doi [\mnras] {10.1093/mnras/staa1419}, 495, 3022

\bibitem[\protect\citeauthoryear{Pace, Walker, Koposov, Caldwell, Mateo,
  Olszewski, III  \& Wang}{Pace et~al.}{2021}]{Pace2021}
Pace A.~B.,  Walker M.~G.,  Koposov S.~E.,  Caldwell N.,  Mateo M.,  Olszewski
  E.~W.,  III J. I.~B.,   Wang M.-Y.,  2021, \mn@doi [\apj]
  {10.3847/1538-4357/ac2cd2}, 923, 77

\bibitem[\protect\citeauthoryear{{Pace}, {Erkal}  \& {Li}}{{Pace}
  et~al.}{2022}]{Pace2022}
{Pace} A.~B.,  {Erkal} D.,   {Li} T.~S.,  2022, arXiv e-prints, \href
  {https://ui.adsabs.harvard.edu/abs/2022arXiv220505699P} {p. arXiv:2205.05699}

\bibitem[\protect\citeauthoryear{Patel, Besla  \& Sohn}{Patel
  et~al.}{2017}]{Patel2017a}
Patel E.,  Besla G.,   Sohn S.~T.,  2017, \mn@doi [\mnras]
  {10.1093/mnras/stw2616}, 464, 3825

\bibitem[\protect\citeauthoryear{Patel, Besla, Mandel  \& Sohn}{Patel
  et~al.}{2018}]{Patel2018}
Patel E.,  Besla G.,  Mandel K.,   Sohn S.~T.,  2018, \mn@doi [\apj]
  {10.3847/1538-4357/aab78f}, 857, 78

\bibitem[\protect\citeauthoryear{Pillepich et~al.,}{Pillepich
  et~al.}{2018a}]{Pillepich2018a}
Pillepich A.,  et~al., 2018a, \mn@doi [\mnras] {10.1093/mnras/stx2656}, 473,
  4077

\bibitem[\protect\citeauthoryear{Pillepich et~al.,}{Pillepich
  et~al.}{2018b}]{Pillepich2018b}
Pillepich A.,  et~al., 2018b, \mn@doi [\mnras] {10.1093/mnras/stx3112}, 475,
  648

\bibitem[\protect\citeauthoryear{Pillepich et~al.,}{Pillepich
  et~al.}{2019}]{Pillepich2019}
Pillepich A.,  et~al., 2019, \mn@doi [\mnras] {10.1093/mnras/stz2338}, 490,
  3196

\bibitem[\protect\citeauthoryear{{Planck Collaboration} et~al.,}{{Planck
  Collaboration} et~al.}{2016}]{PlanckCollaboration2016}
{Planck Collaboration} et~al., 2016, \mn@doi [A\&A]
  {10.1051/0004-6361/201525830}, 594, A13

\bibitem[\protect\citeauthoryear{Putman, Zheng, Price-Whelan, Grcevich,
  Johnson, Tollerud  \& Peek}{Putman et~al.}{2021}]{Putman2021}
Putman M.~E.,  Zheng Y.,  Price-Whelan A.~M.,  Grcevich J.,  Johnson A.~C.,
  Tollerud E.,   Peek J. E.~G.,  2021, \mn@doi [\apj]
  {10.3847/1538-4357/abe391}, 913, 53

\bibitem[\protect\citeauthoryear{Qi, Zivick, Pace, Riley  \& Strigari}{Qi
  et~al.}{2022}]{Qi2022}
Qi Y.,  Zivick P.,  Pace A.~B.,  Riley A.~H.,   Strigari L.~E.,  2022, \mn@doi
  [\mnras] {10.1093/mnras/stac805}, 512, 5601

\bibitem[\protect\citeauthoryear{Reid \& Brunthaler}{Reid \&
  Brunthaler}{2004}]{Reid2004}
Reid M.~J.,  Brunthaler A.,  2004, \mn@doi [\apj] {10.1086/424960}, 616, 872

\bibitem[\protect\citeauthoryear{Riello et~al.,}{Riello
  et~al.}{2021}]{Riello2020}
Riello M.,  et~al., 2021, \mn@doi [A\&A] {10.1051/0004-6361/202039587}, 649, A3

\bibitem[\protect\citeauthoryear{Sohn et~al.,}{Sohn et~al.}{2017}]{Sohn2017}
Sohn S.~T.,  et~al., 2017, \mn@doi [\apj] {10.3847/1538-4357/aa917b}, 849, 93

\bibitem[\protect\citeauthoryear{Springel}{Springel}{2010}]{Springel2010}
Springel V.,  2010, \mn@doi [\mnras] {10.1111/j.1365-2966.2009.15715.x}, 401,
  791

\bibitem[\protect\citeauthoryear{Springel, White, Tormen  \&
  Kauffmann}{Springel et~al.}{2001}]{Springel2001}
Springel V.,  White S. D.~M.,  Tormen G.,   Kauffmann G.,  2001, \mn@doi
  [\mnras] {10.1046/j.1365-8711.2001.04912.x}, 328, 726

\bibitem[\protect\citeauthoryear{Springel et~al.,}{Springel
  et~al.}{2018}]{Springel2018}
Springel V.,  et~al., 2018, \mn@doi [\mnras] {10.1093/mnras/stx3304}, 475, 676

\bibitem[\protect\citeauthoryear{{Taibi} et~al.,}{{Taibi}
  et~al.}{2018}]{Taibi2018}
{Taibi} S.,  et~al., 2018, \mn@doi [\aap] {10.1051/0004-6361/201833414}, \href
  {https://ui.adsabs.harvard.edu/abs/2018A&A...618A.122T} {618, A122}

\bibitem[\protect\citeauthoryear{{Taibi}, {Battaglia}, {Rejkuba}, {Leaman},
  {Kacharov}, {Iorio}, {Jablonka}  \& {Zoccali}}{{Taibi}
  et~al.}{2020}]{Taibi2020}
{Taibi} S.,  {Battaglia} G.,  {Rejkuba} M.,  {Leaman} R.,  {Kacharov} N.,
  {Iorio} G.,  {Jablonka} P.,   {Zoccali} M.,  2020, \mn@doi [\aap]
  {10.1051/0004-6361/201937240}, \href
  {https://ui.adsabs.harvard.edu/abs/2020A&A...635A.152T} {635, A152}

\bibitem[\protect\citeauthoryear{Vitral}{Vitral}{2021}]{Vitral2021}
Vitral E.,  2021, \mn@doi [\mnras] {10.1093/mnras/stab947}, 504, 1355

\bibitem[\protect\citeauthoryear{Walker, Mateo  \& Olszewski}{Walker
  et~al.}{2008}]{Walker2008}
Walker M.~G.,  Mateo M.,   Olszewski E.~W.,  2008, \mn@doi [\apj]
  {10.1086/595586}, 688, L75

\bibitem[\protect\citeauthoryear{Walker, Mateo  \& Olszewski}{Walker
  et~al.}{2009a}]{Walker2009}
Walker M.~G.,  Mateo M.,   Olszewski E.~W.,  2009a, \mn@doi [\aj]
  {10.1088/0004-6256/137/2/3100}, 137, 3100

\bibitem[\protect\citeauthoryear{Walker, Mateo, Olszewski, Sen  \&
  Woodroofe}{Walker et~al.}{2009b}]{Walker2009clean}
Walker M.~G.,  Mateo M.,  Olszewski E.~W.,  Sen B.,   Woodroofe M.,  2009b,
  \mn@doi [\aj] {10.1088/0004-6256/137/2/3109}, 137, 3109

\bibitem[\protect\citeauthoryear{Walker, Mateo, Olszewski, Pe{\~{n}}arrubia,
  Evans  \& Gilmore}{Walker et~al.}{2009c}]{Walker2009univ}
Walker M.~G.,  Mateo M.,  Olszewski E.~W.,  Pe{\~{n}}arrubia J.,  Evans N.~W.,
   Gilmore G.,  2009c, \mn@doi [\apj] {10.1088/0004-637x/704/2/1274}, 704, 1274

\bibitem[\protect\citeauthoryear{Walker, Olszewski  \& Mateo}{Walker
  et~al.}{2015}]{Walker2015}
Walker M.~G.,  Olszewski E.~W.,   Mateo M.,  2015, \mn@doi [\mnras]
  {10.1093/mnras/stv099}, 448, 2717

\bibitem[\protect\citeauthoryear{Weinberger et~al.,}{Weinberger
  et~al.}{2017}]{Weinberger2017}
Weinberger R.,  et~al., 2017, \mn@doi [\mnras] {10.1093/mnras/stw2944}, 465,
  3291

\bibitem[\protect\citeauthoryear{{Weisz}, {Dolphin}, {Skillman}, {Holtzman},
  {Gilbert}, {Dalcanton}  \& {Williams}}{{Weisz} et~al.}{2014}]{Weisz2014}
{Weisz} D.~R.,  {Dolphin} A.~E.,  {Skillman} E.~D.,  {Holtzman} J.,  {Gilbert}
  K.~M.,  {Dalcanton} J.~J.,   {Williams} B.~F.,  2014, \mn@doi [\apj]
  {10.1088/0004-637X/789/2/147}, \href
  {https://ui.adsabs.harvard.edu/abs/2014ApJ...789..147W} {789, 147}

\bibitem[\protect\citeauthoryear{Wheeler et~al.,}{Wheeler
  et~al.}{2017}]{Wheeler2017}
Wheeler C.,  et~al., 2017, \mn@doi [\mnras] {10.1093/mnras/stw2583}, 465, 2420

\bibitem[\protect\citeauthoryear{White \& Rees}{White \&
  Rees}{1978}]{WhiteRees1978}
White S. D.~M.,  Rees M.~J.,  1978, \mn@doi [\mnras] {10.1093/mnras/183.3.341},
  183, 341

\bibitem[\protect\citeauthoryear{Wilkinson, Kleyna, Evans, Gilmore, Irwin  \&
  Grebel}{Wilkinson et~al.}{2004}]{Wilkinson2004}
Wilkinson M.~I.,  Kleyna J.~T.,  Evans N.~W.,  Gilmore G.~F.,  Irwin M.~J.,
  Grebel E.~K.,  2004, \mn@doi [\apj] {10.1086/423619}, 611, L21

\bibitem[\protect\citeauthoryear{Zhu, van~de Ven, Watkins  \& Posti}{Zhu
  et~al.}{2016}]{Zhu2016}
Zhu L.,  van~de Ven G.,  Watkins L.~L.,   Posti L.,  2016, \mn@doi [\mnras]
  {10.1093/mnras/stw2081}, 463, 1117

\bibitem[\protect\citeauthoryear{{de Boer}, {Tolstoy}, {Lemasle}, {Saha},
  {Olszewski}, {Mateo}, {Irwin}  \& {Battaglia}}{{de Boer}
  et~al.}{2014}]{deBoer2014}
{de Boer} T.~J.~L.,  {Tolstoy} E.,  {Lemasle} B.,  {Saha} A.,  {Olszewski}
  E.~W.,  {Mateo} M.,  {Irwin} M.~J.,   {Battaglia} G.,  2014, \mn@doi [\aap]
  {10.1051/0004-6361/201424119}, \href
  {https://ui.adsabs.harvard.edu/abs/2014A&A...572A..10D} {572, A10}

\bibitem[\protect\citeauthoryear{del Pino}{del Pino}{2022}]{GetGaia}
del Pino A.,  2022, AndresdPM/GetGaia: GetGaia,
  \mn@doi{10.5281/zenodo.6473476}, \url
  {https://doi.org/10.5281/zenodo.6473476}

\bibitem[\protect\citeauthoryear{del Pino, Hidalgo, Aparicio, Gallart, Carrera,
  Monelli, Buonanno  \& Marconi}{del Pino et~al.}{2013}]{delPino2013}
del Pino A.,  Hidalgo S.~L.,  Aparicio A.,  Gallart C.,  Carrera R.,  Monelli
  M.,  Buonanno R.,   Marconi G.,  2013, \mn@doi [\mnras]
  {10.1093/mnras/stt833}, 433, 1505

\bibitem[\protect\citeauthoryear{del Pino, Aparicio, Hidalgo  \& {\L}okas}{del
  Pino et~al.}{2017a}]{delPino2017}
del Pino A.,  Aparicio A.,  Hidalgo S.~L.,   {\L}okas E.~L.,  2017a, \mn@doi
  [\mnras] {10.1093/mnras/stw3016}, 465, 3708

\bibitem[\protect\citeauthoryear{del Pino, Łokas, Hidalgo  \& Fouquet}{del
  Pino et~al.}{2017b}]{delPino2017b}
del Pino A.,  Łokas E.~L.,  Hidalgo S.~L.,   Fouquet S.,  2017b, \mn@doi
  [\mnras] {10.1093/mnras/stx1195}, 469, 4999

\bibitem[\protect\citeauthoryear{del Pino, Fardal, van~der Marel, {\L}okas,
  Mateu  \& Sohn}{del Pino et~al.}{2021}]{delPino2021}
del Pino A.,  Fardal M.~A.,  van~der Marel R.~P.,  {\L}okas E.~L.,  Mateu C.,
  Sohn S.~T.,  2021, \mn@doi [\apj] {10.3847/1538-4357/abd5bf}, 908, 244

\bibitem[\protect\citeauthoryear{{del Pino} et~al.,}{{del Pino}
  et~al.}{2022}]{delPino2022}
{del Pino} A.,  et~al., 2022, arXiv e-prints, \href
  {https://ui.adsabs.harvard.edu/abs/2022arXiv220508009D} {p. arXiv:2205.08009}

\bibitem[\protect\citeauthoryear{van~der Marel \& Cioni}{van~der Marel \&
  Cioni}{2001}]{VanderMarel2001}
van~der Marel R.~P.,  Cioni M.-R.~L.,  2001, \mn@doi [\aj] {10.1086/323099},
  122, 1807

\bibitem[\protect\citeauthoryear{van~der Marel, Alves, Hardy  \&
  Suntzeff}{van~der Marel et~al.}{2002}]{VanderMarel2002}
van~der Marel R.~P.,  Alves D.~R.,  Hardy E.,   Suntzeff N.~B.,  2002, \mn@doi
  [\aj] {10.1086/343775}, 124, 2639

\bibitem[\protect\citeauthoryear{van~der Marel, Fardal, Sohn, Patel, Besla, del
  Pino, Sahlmann  \& Watkins}{van~der Marel et~al.}{2019}]{VanderMarel2019}
van~der Marel R.~P.,  Fardal M.~A.,  Sohn S.~T.,  Patel E.,  Besla G.,  del
  Pino A.,  Sahlmann J.,   Watkins L.~L.,  2019, \mn@doi [\apj]
  {10.3847/1538-4357/ab001b}, 872, 24

\makeatother
\end{thebibliography}






\bsp	
\label{lastpage}
\end{document}